\documentclass[a4paper,11pt]{article}
\usepackage{amsmath,amsfonts,amssymb}
\usepackage[nosort]{cite}
\usepackage[top=3.5cm, bottom=3.5cm, right=3cm, left=3cm]{geometry}
\usepackage{hyperref}
\usepackage{graphicx}
\usepackage{color}
\usepackage{epsfig}
\usepackage{multirow}
\usepackage{psfrag}	
\usepackage{tikz}
\usepackage{dsfont}
\usetikzlibrary{snakes}
\usepackage{wrapfig}
\usepackage{slashed}
\usepackage{ulem}
\usepackage[font=small,labelfont=bf,width=1\textwidth]{caption}
\usepackage{subfigure}
\usepackage{comment}
\usepackage[all]{xy}
\usepackage{multirow}
\usepackage{float}
\usepackage{mathtools}
\usepackage{authblk}

\usepackage{color}

\newcommand \be{\begin{eqnarray}}
\newcommand \ee{\end{eqnarray}}

\numberwithin{equation}{section}

\DeclareMathOperator{\llangle}{\big\langle\hspace{-1.2mm}\big\langle\hspace{-.5mm}}
\DeclareMathOperator{\rrangle}{\hspace{-.5mm}\big\rangle\hspace{-1.2mm}\big\rangle}
\DeclareMathOperator{\Tr}{Tr}

\DeclareMathOperator{\sTr}{sTr}

\DeclareMathOperator{\diag}{diag}

\newcommand{\bea}{\begin{eqnarray}}
\newcommand{\eea}{\end{eqnarray}}
\newcommand{\beq}{\begin{equation}}
\newcommand{\eeq}{\end{equation}}
\newcommand{\bal}{\begin{equation}\begin{aligned}}
\newcommand{\eal}{\end{aligned} \end{equation}}

\newcommand{\vev}[1]{{\left< {#1} \right>}}

\newcommand{\cA}{{\mathcal A}}

\newcommand{\cL}{{\mathcal L}}

\newcommand{\cP}{{\mathcal P}}
\newcommand{\cQ}{{\mathcal Q}}
\newcommand{\cT}{{\mathcal T}}
\newcommand{\cR}{{\mathcal R}}
\newcommand{\cO}{{\mathcal O}}

\newcommand{\cI}{{\mathcal I}}

\usepackage{tikz}   
\usepackage{tikz-3dplot}

\title{\bf{Interpolating Bremsstrahlung function in ABJM}}
\author[a]{Luigi Castiglioni\footnote{email: l.castiglioni8@campus.unimib.it}}
\author[a]{Silvia Penati\footnote{email: silvia.penati@mib.infn.it}}
\author[a]{Marcia Tenser\footnote{email: marciatenser@gmail.com}}
\author[b,c]{Diego Trancanelli\footnote{email: dtrancan@gmail.com}}
\affil[a]{\it{Dipartimento di Fisica, Universit\`a degli Studi di Milano--Bicocca and \newline INFN, Sezione di Milano--Bicocca, Piazza della Scienza 3, 20126 Milano, Italy\newline}}
\affil[b]{\it{Dipartimento di Scienze Fisiche, Informatiche e Matematiche, Universit\`a di Modena e Reggio Emilia, via G. Campi 213/A, 41125 Modena, Italy\newline }}
\affil[c]{\it{INFN Sezione di Bologna, via Irnerio 46, 40126 Bologna, Italy}}
\date{}

\begin{document}
\maketitle
\thispagestyle{empty}

\begin{abstract}
\noindent 
    In ABJM theory, enriched RG flows between circular 1/6 BPS bosonic and 1/2 BPS fermionic Wilson loops have been introduced in {\tt arXiv:2211.16501}. These flows are triggered by deformations corresponding to parametric 1/6 BPS fermionic loops. In this paper we revisit the study of these operators, but instead of circular contours we consider an interpolating cusped line and a latitude and study their RG flow in perturbation theory. This allows for the definition of a Bremsstrahlung function away from fixed points. We generalize to this case the known cusp/latitude correspondence that relates the Bremsstrahlung function to a latitude Wilson loop. We find that away from the conformal fixed points the ordinary identity is broken by the conformal anomaly in a controlled way. From a defect perspective,  the breaking of the correspondence can be traced back to the appearance of an anomalous dimension for fermionic operators localized on the defect. As a by-product, we provide a brand new result for the two-loop cusp anomalous dimension of the 1/6 BPS fermionic and the 1/6 BPS bosonic Wilson lines. 
\end{abstract}

\newpage 

\tableofcontents


\setcounter{page}{1}

\section{Introduction \& results}

Wilson loop (WL) operators can be seen as dynamical one-dimensional defects embedded in higher dimensional quantum field theories  \cite{Polchinski:2011im,Cooke:2017qgm,Giombi:2017cqn,Beccaria:2017rbe,Giombi:2018qox, Liendo:2018ukf,Beccaria:2019dws,Bianchi:2020hsz,Agmon:2020pde,Beccaria:2022bcr,Drukker:2022txy}. In superconformal settings, besides the conformal generators acting along their contour, they may also preserve a fraction of supersymmetries of the bulk theory. This happens if the usual gauge holonomy is replaced by the holonomy of a generalized connection that includes also couplings to matter fields. Notable examples are the BPS WLs in $\mathcal{N}=4$ SYM theory in four dimensions (see, for example, \cite{Maldacena_1998,erickson,gross,Zarembo:2002an,Drukker:2007qr}) and in ABJ(M) theory in three dimensions (see, for example, \cite{Drukker:2008zx,Chen:2008bp, Drukker:2009hy,Cardinali:2012ru,Bianchi:2014laa,Ouyang:2015iza} and \cite{Drukker:2019bev} for a rather recent review). The contours on which these BPS WLs are supported might be an infinite line or a circle or even a more general curve, as for example latitudes of a sphere or cusps, which are the cases we are considering in this paper.

Near a cusp, the vacuum expectation value (VEV) of WLs acquires short distance divergences. Being  $\varphi$ the cusp angle, in four dimensions the non-Abelian exponentiation theorem \cite{Dotsenko:1979wb, Gatheral:1983cz, Frenkel:1984pz} ensures that the VEV takes the universal form 
\begin{equation}
\label{eq:exponentiation}
    \langle W_{\text{cusp}}\rangle \sim \exp\left(-\Gamma_{\rm cusp}(\varphi) \log \frac{\Lambda}{\mu} + \mbox{finite}\right)\,,
\end{equation}
where $\Gamma_{\rm cusp}(\varphi)$ is the so-called \emph{cusp anomalous dimension} and $\Lambda,\mu$ are the infrared and ultraviolet cutoffs, respectively. In three dimensions an analogue exponentiation theorem is not known. However, for 1/2 BPS WLs exponentiation does seem to occur \cite{Griguolo:2012iq, Bianchi:2017svd}. 

In the small angle limit, $\Gamma_{\rm cusp}(\varphi)$ is governed by the \emph{Bremsstrahlung function} $B^\varphi$, according to
\begin{equation}\label{eq:Bdefinition}
    \Gamma_{\rm cusp}(\varphi)\underset{\varphi \ll 1}{\sim} -\varphi^2 B^\varphi\,.
\end{equation}
Its name follows from the fact that in conformal field theories $B^{\varphi}$ is related to the energy radiated by a very massive particle accelerating in a gauge background, and it can be associated with the coefficient of the two-point function of the displacement operator \cite{Correa:2012at}. 

In addition to the \emph{geometrical cusp} parameterized by $\varphi$, it is also possible to introduce an \emph{internal cusp} angle $\theta$, appearing in the operator definition through the coupling to matter and describing the change of orientation in the R-symmetry space. In this case the cusp anomalous dimension - now called {\em generalized cusp anomalous dimension} - will acquire a non-trivial dependence also on $\theta$. Taking the small $\varphi, \theta$ angles limit, one then defines two (generically different) Bremsstrahlung functions, as
\begin{equation}\label{eq:ThetaPhiintro}
    \Gamma_{\text{cusp}}(\varphi, \theta) \sim \theta^2 B^{\theta} - \varphi^2 B^{\varphi}\,.
\end{equation}
Similarly to the geometrical case, $B^{\theta}$ can be associated with the two-point function of an R-symmetry displacement operator \cite{Aguilera-Damia:2014bqa}.

A particular feature of ABJ(M) theory is that there is a rich spectrum of supersymmetric WLs (for a review, see \cite{Drukker:2019bev}). For some of the representatives, results regarding the corresponding Bremsstrahlung functions are already known. Notable examples are the $1/2$ fermionic BPS operator \cite{Drukker:2009hy} and the 1/6 BPS bosonic one \cite{Drukker:2008zx}, defined on the maximal circle on the 3-sphere. In the 1/2 BPS case $\Gamma_{\rm cusp}$ becomes supersymmetric at $\theta = \varphi$, implying that $B^{\theta}=B^{\varphi} (\equiv B_{1/2})$. The two-loop computation of $B_{1/2}$ was done in \cite{Griguolo:2012iq} and its three-loop value was found in \cite{Bianchi:2017svd}.
For the bosonic $1/6$ BPS operator the two $B$-functions are instead related by
$ B^{\theta}= \tfrac12 B^{\varphi} (\equiv B_{1/6})$ \cite{Bianchi:2017afp, Bianchi:2018scb}, as a consequence of superconformal Ward identities of the bulk theory.
The $B_{1/6}$ function has been determined up to four loops \cite{Bianchi:2017afp,Bianchi:2017ujp}. 

In both cases an exact formula has been proposed in \cite{Bianchi:2014laa} and later proved in \cite{Correa:2014aga, Bianchi:2017ozk}, which relates the Bremsstrahlung function 
to the derivative of a {\em latitude} WL on the 3-sphere with respect to the latitude angle, in analogy with what happens for ${\cal N}=4$ SYM in four dimensions \cite{Correa:2012at}. Since the BPS latitude WL is in principle amenable to exact evaluation via a matrix integral \cite{Bianchi:2018bke,Griguolo:2021rke}, this identity opens the possibility of computing the Bremsstrahlung function exactly.   
On the other hand, in the 1/2 BPS case the exact evaluation of $\Gamma_{\rm cusp}$, and consequently of $B_{1/2}$, can be formulated as a boundary Thermodynamic Bethe Ansatz that leads to a $Y$-system of integrable equations \cite{Correa:2023lsm}. A comparison between the localization-based evaluation of $B_{1/2}$ and the one obtained by solving the $Y$-system would allow for an explicit determination of the ubiquitous $h(\lambda)$ function of ABJ(M), whose explicit expression has been conjectured in \cite{Gromov:2014eha}. 

Given the remarkable role played by the Bremsstrahlung function in understading the relation between superconformal invariance and integrability in ABJ(M) theory, it is important to generalize our investigation to other classes of WL representatives. 

Realizations of Wilson loop operators interpolating between $1/2$ and $1/6$ BPS representatives were introduced in \cite{Drukker:2019bev} and can be obtained following the ``hyperloop prescription'' \cite{Drukker:2020dvr}. As a result, a new class of {\em parametric} WLs can be defined,  which are still supported along the maximal circle on the 3-sphere but depend on 
complex constant parameters $\bar\alpha$ and $\alpha$ continuously interpolating between $1/6$ BPS bosonic ($\bar\alpha\alpha=0$) and $1/2$ BPS ($\bar\alpha\alpha=1$) WLs, where all other points in the parameter space correspond to $1/6$ BPS fermionic operators. These operators undergo a non-trivial renormalization \cite{Castiglioni:2022yes,Castiglioni:2023uus} that can be assessed via a suitable generalization of the one-dimensional auxiliary method \cite{Samuel:1978iy, Gervais:1979fv, Arefeva:1980zd, CRAIGIE1981204, Aoyama:1981ev,   Dorn:1986dt} originally introduced
in the context of QCD. Renormalization group flows connecting these representatives, referred to as enriched RG flows, are such that $1/6$ BPS bosonic as well as $1/2$ BPS operators sit at fixed points. From a defect perspective, 1/6 BPS bosonic and 1/2 BPS WLs support superconformal one-dimensional theories, while 1/6 BPS fermionic ones give rise to defects that are still supersymmetric but no longer scale and conformally invariant. 

\begin{figure}[h!]
    \centering
    \includegraphics[width=.95\textwidth]{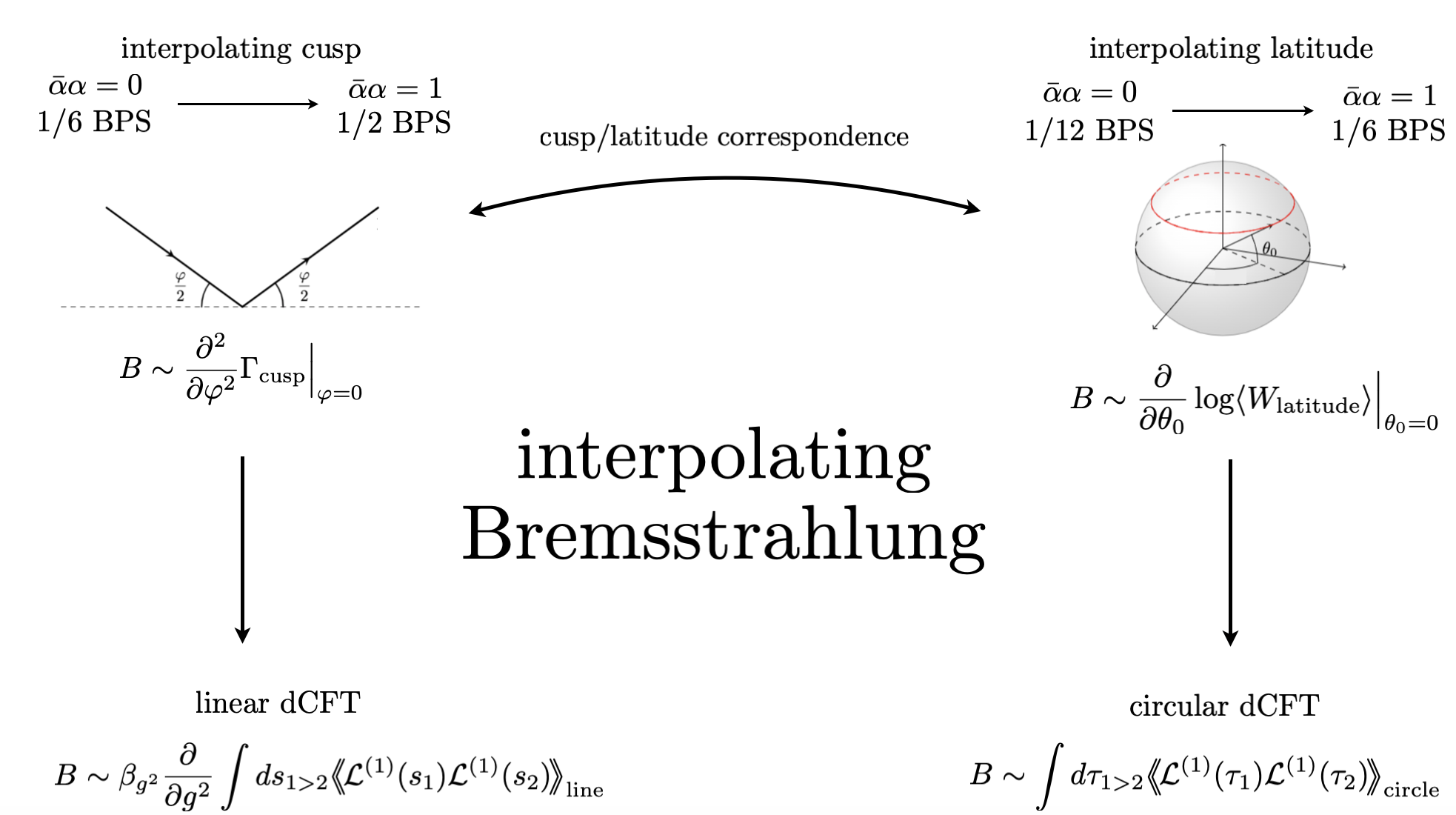}
    \caption{The main quantity defined in this paper, the interpolating Bremsstrahlung function, has different avatars and can be computed in different ways. (top left) From the interpolating cusp studied in Sec. \ref{sec:Cusp}, it can be obtained by taking derivatives of $\Gamma_\textrm{cusp}$ with respect to the cusp angle $\varphi$. For $\varphi\neq 0$ this operator does not preserve any supersymmetry, however we indicate what would be the supersymmetries preserved by the fixed points in the zero cusp limit, depending on the value of $\bar\alpha\alpha$. (top right) The interpolating latitude, that flows between a 1/12 BPS operator for $\bar\alpha\alpha=0$ and a 1/6 BPS one for $\bar\alpha\alpha=1$, is studied in Sec. \ref{sec:latitudeWLs} and gives the interpolating Bremsstrahlung by taking a derivative with respect to the latitude angle. These computations are related by a cusp/latitude correspondence studied at the end of Sec. \ref{sec:defectcorrelationfunction}. (bottom) In that section, we also report the computation of the interpolating Bremsstrahlung from a defect CFT point of view, i.e. from the integrated 2-point function on the Wilson loop of the term linear in $\theta$ of the superconnection ${\cal L}^{(1)}$.}
    \label{fig1}
\end{figure}

We now proceed with a summary of our results. 
A schematic map of the relation among the computations in this paper can be found in Fig. \ref{fig1}.

\paragraph{The interpolating generalized cusp.} Here we put together the original construction of the generalized cusped Wilson line \cite{Griguolo:2012iq} (see also \cite{Drukker:2011za} for the original construction in ${\cal N}=4$ super Yang-Mills theory) and the hyperloop tools \cite{Drukker:2020dvr} to construct cusp deformations of parametric WLs. The result is an \emph{interpolating generalized cusp} connecting $1/2$ and $1/6$ cusp representatives, which exhibits a non-trivial dependence on the $\bar\alpha, \alpha$ parameters in addition to the angular functional dependence. 
We investigate this operator at quantum level 
using the one-dimensional auxiliary method  \cite{Castiglioni:2022yes}. In this set-up the corresponding \emph{interpolating cusp anomalous dimension} 
$\Gamma_{\text{cusp}}(\varphi, \theta, \bar\alpha, \alpha)$ can be inferred from the renormalization function of the composite one-dimensional operator localized at the cusp. The remarkable result is that away from the two fixed points short distance divergences no longer exponentiate and we cannot rely on equation \eqref{eq:exponentiation} to determine $\Gamma_{\text{cusp}}(\varphi, \theta, \bar\alpha, \alpha)$. Alternatively, we apply the standard definition of the anomalous dimension as the derivative of the renormalization function with respect to the renormalization scale. At two loops in the ABJM coupling constant $N/k$, where $N$ is the rank of the two gauge groups and $k$ is the Chern-Simons level\footnote{We work in the large $N$ limit, with $\tfrac{N}{k} \ll 1$.},  we find
\begin{align}   \label{eq:cuspandimintro}
\Gamma_{\text{cusp}}(\varphi, \theta, \bar\alpha, \alpha) =&\frac{N}{k}\bar\alpha\alpha \left( 1-\frac{\cos\frac{\theta}{2}}{\cos\frac{\varphi}{2}} \right) \\
&
+
    \frac{N^2}{k^2}\Bigg[ \left[(1-\bar\alpha^2\alpha^2)\left(2\cos\varphi - \cos\theta \right) -(\bar\alpha\alpha-1)^2 \right]\frac{\varphi}{2\sin\varphi} \nonumber \\ &\hskip 1.8cm +\bar\alpha\alpha(\bar\alpha\alpha-1)+2\bar\alpha^2\alpha^2\left(\frac{\cos\frac{\theta}{2}}{\cos\frac{\varphi}{2}} -1 \right)\log\sec\frac{\varphi}{2} \nonumber \\
    & \hskip 1.8cm 
    +\bar\alpha\alpha(\bar\alpha\alpha-1)\left( \frac{\cos\frac{\theta}{2}}{\cos\frac{\varphi}{2}} - 1 \right)\log\left( \frac{c_J}{c_z} \mu^2L^2 \right)\Bigg]
    + \cO \left( \frac{N^3}{k^3} \right) \,. \nonumber 
\end{align}

The result is generically scheme dependent, similarly to what has been observed in four-dimensional theories \cite{Beccaria:2018ocq,Billo:2019job, Billo:2023igr}. Scheme dependence is enclosed in the last term of the formula above into the constant $c_J$, which is associated to the renormalization of a composite operator localized at the cusp, and the constant $c_z$, which originates from the renormalization of the one-dimensional auxiliary field. Moreover, the logarithm also contains the renormalization scale $\mu$ and an IR cutoff $L$ introduced in the procedure. We shall discuss this term in detail. 

For $\bar\alpha\alpha=1$, expression \eqref{eq:cuspandimintro} appropriately reproduces the result of \cite{Griguolo:2012iq}. For $\bar\alpha,\alpha=0$, it gives the cusp anomalous dimension of the cusped $1/6$ BPS bosonic operator,
\begin{equation}
    \Gamma_{\rm cusp}^{1/6} = \frac{N^2}{k^2} \frac{\varphi}{2\sin\varphi} (2\cos\varphi-\cos\theta-1) +  \cO \left( \frac{N^3}{k^3} \right)\,,
\end{equation}
which, to our knowledge, has never been computed before. For generic values of $\bar\alpha, \alpha$ expression
\eqref{eq:cuspandimintro} provides a brand new result for the cusp anomalous dimension of 1/6 BPS fermionic WLs. 

\paragraph{The interpolating Bremsstrahlung functions.} Mimicking the definitions \eqref{eq:ThetaPhiintro} we can now define $\varphi$ and $\theta$ {\em interpolating Bremsstrahlung functions} by expanding \eqref{eq:cuspandimintro} in the small angles limit, thus obtaining (up to order $N^2/k^2$)
\begin{equation}
\label{eq:interpolatingB}
\begin{split}
    B^{\theta} (\bar\alpha\alpha)&= \frac{1}{8}\frac{N}{k} \bar\alpha\alpha -  \frac{1}{4}\frac{N^2}{ k^2} (\bar\alpha^2\alpha^2-1) - \frac{1}{16}\frac{N}{ k} \beta_{\bar\alpha\alpha}\log\left(\frac{c_J}{c_z} \mu^2 L^2 \right) \,,\\
    B^{\varphi} (\bar\alpha\alpha) &= \frac{1}{8}\frac{N}{k} \bar\alpha\alpha -  \frac{1}{2}\frac{N^2}{k^2} (\bar\alpha^2\alpha^2-1) - \frac{1}{16}\frac{N}{k}\beta_{\bar\alpha\alpha}\log\left( 
    \frac{c_J}{c_z}  \mu^2 L^2 e^{-\frac{4}{3}} \right)  \,.
\end{split}
\end{equation}

For generic values of the parameters, they are the Bremsstrahlung functions of the 1/6 BPS fermionic WLs. They correctly interpolate between the two-loop values of the 1/6 BPS bosonic and 1/2 BPS Bremsstrahlung functions. Away from the two fixed points, the corresponding known identities $B^{\varphi} = 2 B^{\theta}$ (for $\bar\alpha \alpha=0$) and $B^{\varphi} =  B^{\theta}$ (for $\bar\alpha \alpha=1$) are broken by additive terms proportional to $\beta_{\bar\alpha \alpha}$, the parameter $\beta$-function which drives the RG flow. 
The particular parametric dependence of the one- and two-loop results clarifies the mechanism which makes $B_{1/6}$ an even function of the $N/k$ coupling, while $B_{1/2}$ is  odd. Along the RG trajectories they do not possess any specific parity. 

\paragraph{The interpolating cusp/latitude correspondence.} An interesting question is whether the interpolating Bremmstrahlung functions \eqref{eq:interpolatingB} are  related to some parametric latitude WLs, as it occurs at the two superconformal fixed points. In principle we do not expect that, since the identities relating cusped and latitude WLs (see \eqref{eq:generalidentity} below)  heavily rely on superconformal invariance, which is broken by the parameter renormalization. Nevertheless it is interesting to understand the breaking mechanism.  

The best candidate entering a parametric deformation of the cusp/latitude correspondence is the $\bar\alpha, \alpha$ {\em parametric latitude} constructed in \cite{Castiglioni:2022yes} and revisited in section \ref{sec:latitudeWLs}. 
It interpolates between two fixed points that correspond to the latitude deformations of the 1/6 BPS bosonic ($\bar\alpha, \alpha =0$) and 1/2 BPS ($\bar\alpha \alpha =1$) operators defined on the maximal circle. A perturbative evaluation of its VEV reveals that away from the two fixed points the result is divergent at two loops. However, renormalization of the parameters appropriately removes UV divergences. These divergences 
vanish at the two fixed points, in agreement with previous results \cite{Bianchi:2014laa}. They also vanish at 
zero latitude angle, consistently with the finiteness of the two-loop result found in the case of the parametric WL defined on the  maximal circle \cite{Castiglioni:2022yes}.
 For a parametric WL  defined on a latitude contour featured by radius $R$ and angle $\theta_0$ ($\cos{\theta_0} \equiv \nu $) our finding at two loops and at renormalization scale $\mu$ is
\begin{align}
    \langle W_{\nu} \rangle & = 1- \frac{N}{k} \pi \bar\alpha\alpha \, \nu \cot\frac{\pi\nu}{2}  - \frac{N^2}{k^2} \frac{\pi}{6} \left( 
3\nu^2(\bar\alpha^2\alpha^2-1) -2\right)\\&+  \frac{N}{k} \beta_{\bar\alpha \alpha}  \left[ \pi^2 \nu+ \frac{\pi}{2} \left(\log\left( \frac{4\pi e^{\gamma_E}\nu^2}{c_z} \mu^2R^2\right)+H_{-\frac{1+\nu}{2}}-3H_{\frac{\nu-1}{2}}\right) \nu \cot\frac{\pi\nu}{2}  \right]  + \cO \left( \frac{N^3}{k^3} \right)\, . \nonumber 
\end{align}
Here $\beta_{\bar\alpha\alpha}$ is the one-loop conformal anomaly associated to the running parameter $\bar\alpha\alpha$. The constant $c_z$ signals scheme dependence, which generically appears already at leading order. 

The comparison between the Bremsstrahlung functions inferred from the cusped WL and the derivative of $\log{\langle W_\nu \rangle}$ with respect to the latitude parameter $\nu$ leads eventually to (up to two loops)
\begin{equation}\label{eq:introB}
 B^{\theta}(\bar\alpha\alpha) =  \frac{1}{4\pi^2} \frac{\partial \log \langle 
    W_{\nu} \rangle}{\partial \nu}\bigg|_{\nu=1} +    
    \beta_{\bar\alpha\alpha}  \,  \frac{1}{16}\frac{N}{k}  \log\left(\frac{4\pi  e^{\gamma_E-2}}{c_J} \mu^2 R^2\right) \,.
\end{equation}
A similar result occurs also for $B^{\varphi}(\bar\alpha\alpha)$, as it differs from $B^{\theta}(\bar\alpha\alpha)$ by $\beta_{\bar\alpha\alpha}$-terms.
At the fixed points we recover the well known cusp/latitude correspondence recalled in \eqref{eq:generalidentity}.  For generic $\alpha\bar\alpha$, instead, such a correspondence is broken by contributions proportional to the $\beta$-function. Non-conformality also introduces scheme dependence. 

\paragraph{Defect interpretation.} (Super)conformal WLs describe one-dimensional defects and integrated correlation functions of certain local operators localized on them are obtained by taking multiple derivatives of the WLs with respect to a parameter, say a latitude or a cusp angle.\footnote{Recently, a formula for the Bremsstrahlung function based on multiple derivatives with respect to masses that can be turned on in the bulk theory has been proposed in \cite{Guerrini:2023rdw}.}
For $1/2$ BPS \cite{Correa:2014aga} and $1/6$ BPS bosonic \cite{Bianchi:2018scb} representatives, such defect correlators can be used to study the Bremsstrahlung function. Here we extend this approach to representatives that do not sit at fixed points, thus deriving $B^\theta$ perturbatively from yet a third point of view. We find remarkable agreement with the expression \eqref{eq:interpolatingB} obtained from the cusp anomalous dimension. 

More generally, we  study defect correlation functions of ABJM operators entering the definition of the defect. We show that the fermionic operators, named $\chi$ below, acquire an anomalous dimension
\begin{equation}
\gamma_\chi=\frac{N}{k}(\bar\alpha\alpha-1) + \cO \left( \frac{N^2}{k^2} \right)\, .
\end{equation}

Furthermore, we reconsider the interpolating cusp/latitude correspondence to better understand in this set-up the mechanism that modifies it away from the fixed points of the RG flow. We find that the deviation from the usual identity valid at the superconformal points can be imputed to  the emergence of the anomalous dimension $\gamma_\chi$ for the fermions localized on the defect. In fact, pertubative investigation up to two loops leads to 
\begin{equation}
    B^{\theta}(\bar\alpha\alpha) = \left[ 1 -\gamma_\chi \log\left(e^2\frac{4L^2}{R^2}\right) \right]\dfrac{1}{4\pi^2}\dfrac{\partial}{\partial\nu}\log \vev{W_\nu}\bigg|_{\nu=1} \,.
\end{equation}
As we discuss later, this result coincides with identity \eqref{eq:introB} when we choose the particular scheme $c_J=16\pi e^{\gamma_E}$.

As a final remark, we stress that our perturbative computations are performed at framing zero.\footnote{This is what is usually done in the ABJM literature, see for example \cite{Bianchi:2013zda, Bianchi:2013rma,Griguolo:2013sma}.} It would be important to repeat them at framing 1, which is what one obtains from supersymmetric localization, following, for example, what has been done in \cite{Bianchi:2016yzj,Bianchi_2016, Bianchi:2018bke} for the 1/6 BPS bosonic Wilson loop and in \cite{Gabai:2022mya} for mesonic Wilson loops. The main problem in our case would be the evaluation of the fermionic Feynman diagrams at framing 1, which is technically challenging.

\vskip 10pt

The rest of the paper is organized as follows. We devote section \ref{app:1dmethod} to a review of the one-dimensional auxiliary theory method and sections \ref{sec:Cusp} and \ref{sec:latitudeWLs} to the cusp and latitude operators, respectively. We collect the results in section \ref{sec:interpolatingBfunction}, where we study the associated interpolating Bremsstrahlung functions and the cusp/latitude correspondence. In section \ref{sec:defectcorrelationfunction} we further explore the latitude WL to study correlation functions of operators defined on the one-dimensional defect. Conventions and Feynman rules are collected in appendix \ref{app:abjm}. For the sake of simplicity we will focus on the $U(N) \times U(N)$ ABJM theory. The generalization to the case of different group ranks is straightforward.
\section{Review of the one-dimensional auxiliary theory method}
\label{app:1dmethod}

We begin by reviewing the one-dimensional auxiliary field method developed in \cite{Samuel:1978iy, Gervais:1979fv, Arefeva:1980zd, CRAIGIE1981204, Aoyama:1981ev,   Dorn:1986dt} to compute Wilson loop VEV in Yang-Mills theories, and generalized in \cite{Castiglioni:2022yes} to  BPS WLs in ABJM theory. 

The one-dimensional auxiliary theory is defined by the effective action
\begin{equation} 
\label{eqn:effS}
\begin{split}
   S_\textrm{eff} = S_\textrm{ABJM} +\int_{\tau_0}^{\tau_1} d\tau\,  \Tr \left(\bar\Psi \left( 
\partial_{\tau} + i\cL \right)\Psi \right)\,,
\end{split}
\end{equation}
where $\cL$ is the superconnection present in the definition of the Wilson loop under investigation and the $\tau$-integral is taken along the Wilson contour. The one-dimensional Grassmann-odd superfield $\Psi$,
\begin{equation}\label{eq:oddmatrix}
    \Psi = \begin{pmatrix} z & \varphi \\ \tilde \varphi& \tilde z\end{pmatrix}\,, \qquad \bar\Psi = \begin{pmatrix} \bar z & \bar{\tilde\varphi} \\ \bar{\varphi}& \bar{\tilde z}\end{pmatrix}\,.
\end{equation}
has components $(z, \tilde z$) and $(\varphi, \tilde \varphi$) that are spinors and  scalars, respectively, in the fundamental representation of $U(N)$. 
The Wilson loop VEV is given by
\begin{equation}
    \langle W(\tau_0,\tau_1) \rangle = \Tr\langle \Psi_0(\tau_1)\bar\Psi_0(\tau_0)\rangle \, .
\end{equation}
The one-dimensional action can be explicitly written as
\begin{equation}
\label{eqn:effectiveaction2node}
   S_{\rm eff} = S_{\rm ABJM} +\int_{\tau_0}^{\tau_1} d\tau \Big[ \bar\varphi D_{\tau}\varphi + \bar{\tilde\varphi} \hat D_{\tau} \tilde \varphi  + \bar z D_{\tau} z + \bar{\tilde z} \hat D_{\tau} \tilde z + i\big(\bar{\tilde z} f \varphi + \bar\varphi \bar f \tilde z + \bar{\tilde\varphi} f z+\bar z\bar f \tilde \varphi\big)\Big]\,,
\end{equation}
where $D_{\tau}\equiv\partial_{\tau} +i\cA$ and $\hat D_{\tau}\equiv\partial_{\tau}+i\hat \cA$ include the generalized connections defined in \eqref{eq:lineWL0}. They give rise to the usual minimal coupling between the one-dimensional fields and the bulk gauge vectors, plus quartic interactions with bulk scalar bilinears through the scalar coupling matrix $M_J^{\ I}$ appearing in $\cL$. 

Each one-dimensional field $\phi=\{\varphi,\tilde{\varphi},z,\tilde{z}\}$ has a corresponding renormalization function $\phi=Z^{-1/2}_\phi\phi_0$,
where $\phi_0$ stands for the bare quantity. Being the action \eqref{eqn:effectiveaction2node} invariant under the formal exchanges $z \leftrightarrow \varphi$ and $\tilde{z} \leftrightarrow \tilde\varphi$, and the exchange of untilde and tilde fields we have $Z_z = Z_\varphi = Z_{\tilde{z}} = Z_{\tilde\varphi}$. In general, both $M_J^{\ I}$ and $f, \bar{f}$ depend on the ABJM coupling as well as on $\alpha, \bar\alpha$ parameters. Since $g^2 \equiv 2\pi/k$ does not renormalize, for the renormalization of the fermionic interactions we define
\begin{equation}
\begin{split}\label{eq:parren}
   & \bar\alpha_0 \, Z^{1/2}_{\tilde z}Z_{\varphi}^{1/2} =\bar\alpha_0 \, Z^{1/2}_{z}Z_{\tilde\varphi}^{1/2}= Z_{\bar\alpha} \, \bar\alpha\,,\\
   & \alpha_0 \,  Z^{1/2}_{\tilde z}Z_{\varphi}^{1/2} =\alpha_0 \, Z^{1/2}_{z}Z_{\tilde\varphi}^{1/2}= Z_{\alpha} \, \alpha\,,
\end{split}
\end{equation}
while for the scalar vertices we set 
\begin{equation}
\begin{alignedat}{2}
\label{eq:Mren}
    &Z_{\varphi}({M}_I^{\ J})_0 =Z_{\varphi C} M_I^{\ J}\,, \qquad &&Z_{\tilde\varphi}({M}_I^{\ J})_0=Z_{\tilde\varphi C} M_I^{\ J}\,,\\ &Z_{z}({M}_I^{\ J})_0=Z_{z C} M_I^{\ J}\,, \qquad &&Z_{\tilde z}\,({M}_I^{\ J})_0=Z_{\tilde z C}M_I^{\ J}\,,
\end{alignedat}
\end{equation}
where $({M}_I^{\ J})_0$ is the scalar coupling matrix expressed in terms of the bare parameters.

We then use the standard BPHZ renormalization procedure and write all renormalization functions as $Z = 1 + \delta$, where $\delta$ are the corresponding countertems. We extract the Feynman rules from the one-dimensional Lagrangian written as the sum of the renormalized Lagrangian plus the counterterm part
\begin{equation}\label{eq:Ltot}
    \cL_\textrm{1D}=\cL^\textrm{ren}_\textrm{1D} + \cL_\textrm{1D}^{\text{ct}}\,.
\end{equation}
The tree-level propagators of the one-dimensional fields are
\begin{equation}
\begin{alignedat}{3}
    & \includegraphics[width=0.2\textwidth]{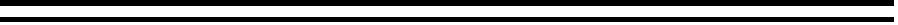} &&= \langle z^i(\tau_1)\bar z_j(\tau_2) \rangle &&= \delta_j^i \, \theta(\tau_1-\tau_2)\,, \\
    & \includegraphics[width=0.2\textwidth]{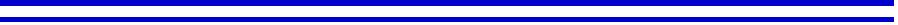} &&= \langle {\tilde z}^{\hat i}(\tau_1)\bar{\tilde z}_{\hat j}(\tau_2) \rangle &&= \delta_{\hat j}^{\hat i} \,  \theta(\tau_1-\tau_2)\,, \\
    & \includegraphics[width=0.2\textwidth]{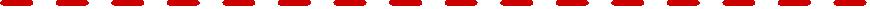} &&= \langle \varphi^i(\tau_1)\bar{\varphi}_j(\tau_2) \rangle &&= \delta_j^i \,  \theta(\tau_1-\tau_2)\,, \\
    & \includegraphics[width=0.2\textwidth]{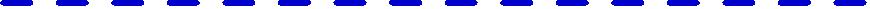} &&= \langle \tilde \varphi^{\hat i}(\tau_1)\bar{\tilde \varphi}_{\hat j}(\tau_2) \rangle &&= \delta_{\hat j}^{\hat i} \,  \theta(\tau_1-\tau_2)\,.
\end{alignedat}
\end{equation}
Following the same prescription adopted in \cite{Castiglioni:2022yes}, we work in the large $N$ limit and study the UV behavior of the one-dimensional theory in the $\tau_2\to\tau_1$ limit by taming short distance divergences arising from the evaluation of Feynman integrals by  
the use  of dimensional regularization in $D = 3-2\epsilon$.

\vskip 5pt


\subsection{Renormalization of the one-dimensional field function}
\label{app:Zz}

In this section we report the highlights of the two-loop computation of the $z$ field renormalization function.

The one-loop result is given by a single diagram that was already computed in \cite{Castiglioni:2022yes}. Here we generalize it to include a renormalization scheme parameter $c_z$, such that
\begin{equation}
    \vcenter{\hbox{\includegraphics[width=0.15\textwidth]{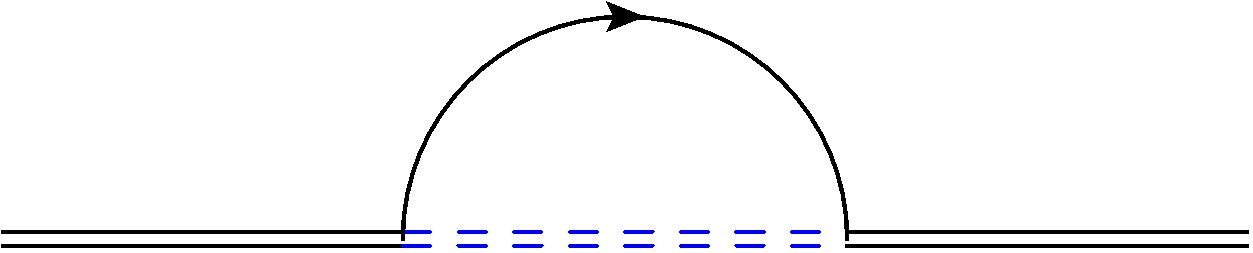}}}  =  -\bar\alpha\alpha \frac{g^2 N}{4\pi}\frac{(c_z)^{\epsilon}}{\epsilon} \int ds \ \bar z \dot z  \,.
\end{equation}
In \cite{Castiglioni:2022yes,Castiglioni:2023uus} we chose implicitly $c_z=1$, which corresponds to MS scheme. Alternatively, one could use a $\overline{\text{MS}}$ scheme, a convenient choice being
$c_z = 4\pi e^{\gamma_E}$. However, we prefer to stick to a more general scheme to better understand how scheme dependence percolates into the results.

At two loops we have to consider the following one-particle irreducible diagrams
\begin{equation}
\begin{split}
    \vcenter{\hbox{\includegraphics[width=0.15\textwidth]{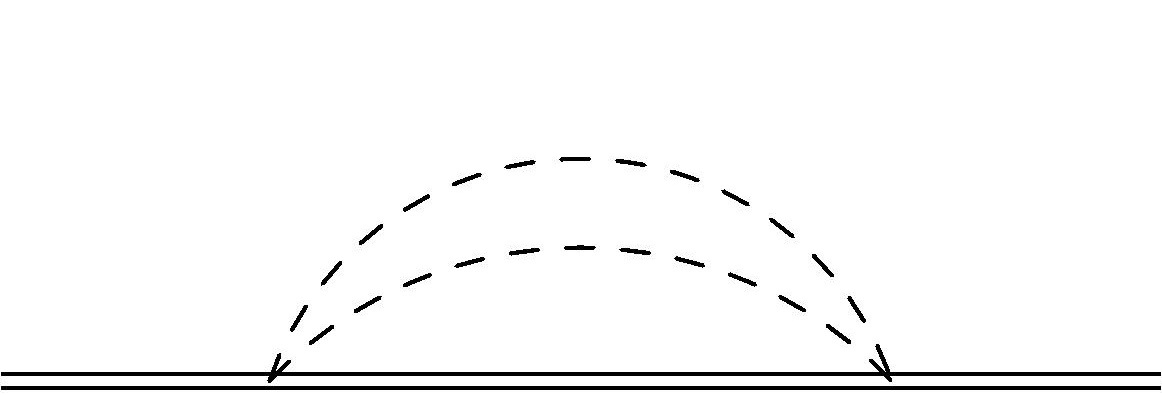}}} &+ \vcenter{\hbox{\includegraphics[width=0.15\textwidth]{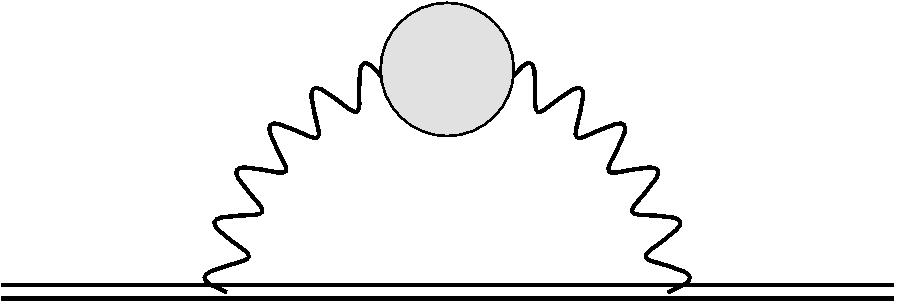}}} =  -\frac{g^4 N^2}{16\pi^2}\alpha\bar\alpha(\alpha\bar\alpha-1)(c_z)^{2\epsilon} \, \frac{1}{\epsilon} \, \int ds \ \bar z \dot z  \,, \nonumber
\end{split}
\end{equation}
    \begin{equation}
\begin{split}
    \vcenter{\hbox{\includegraphics[width=0.15\textwidth]{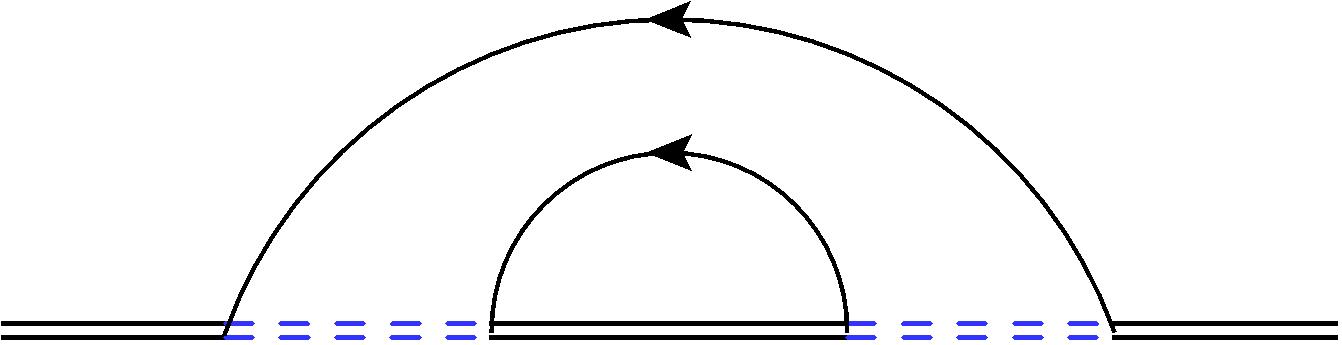}}} &=  \bar\alpha^2\alpha^2 \frac{g^4N^2}{16\pi^2}(c_z)^{2\epsilon} \, \frac{1}{\epsilon} \,\left[ \frac{1}{2\epsilon} + \log\left( 4\pi e^{\gamma_E-2} \right) \right]\int ds \ \bar z \dot z \,,\nonumber
\end{split}
\end{equation}
\begin{equation}
\begin{split}
    \qquad\quad\vcenter{\hbox{\includegraphics[width=0.15\textwidth]{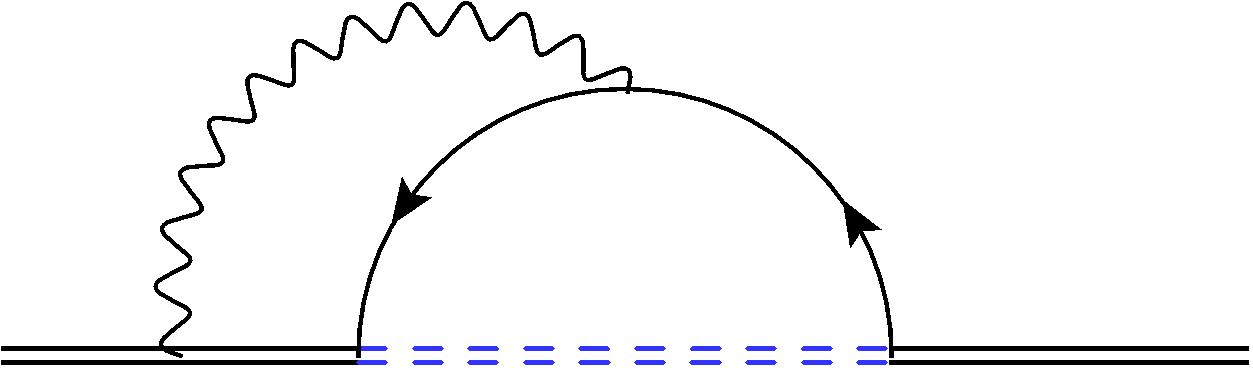}}} &=  -\bar\alpha\alpha \frac{g^4N^2}{32\pi^2}(c_z)^{2\epsilon} \, \frac{1}{\epsilon} \, \left[ \frac{1}{2\epsilon} + \log\left( 4\pi e^{\gamma_E-2} \right)  \right]  \int ds \ \bar z \dot z  \,,
\end{split}
\end{equation}
\begin{equation}
\begin{split}
    \vcenter{\hbox{\includegraphics[width=0.15\textwidth]{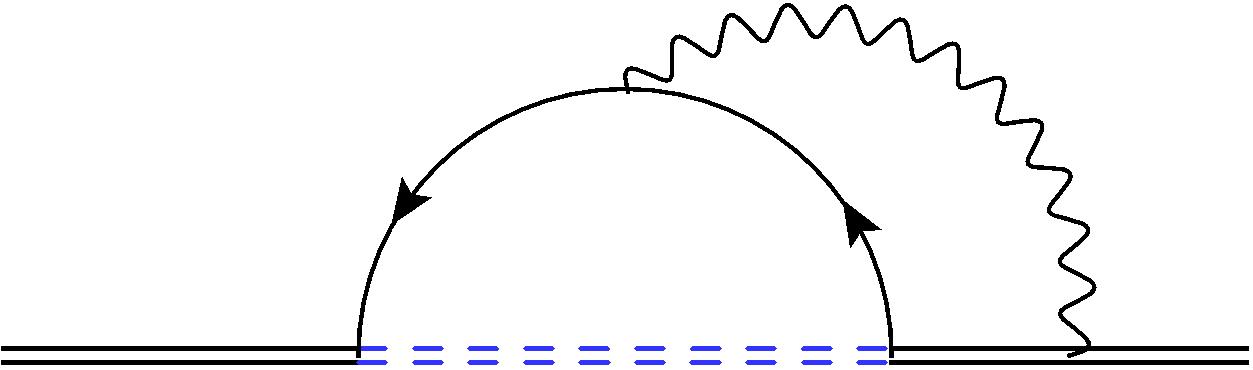}}} &=  -\bar\alpha\alpha \frac{g^4N^2}{32\pi^2}(c_z)^{2\epsilon} \, \frac{1}{\epsilon} \, \left[ \frac{1}{2\epsilon} + \log\left( 4\pi e^{\gamma_E-2} \right)  \right]  \int ds \ \bar z \dot z  \,,\nonumber
\end{split}
\end{equation}
\begin{equation}
\begin{split}
    \vcenter{\hbox{\includegraphics[width=0.15\textwidth]{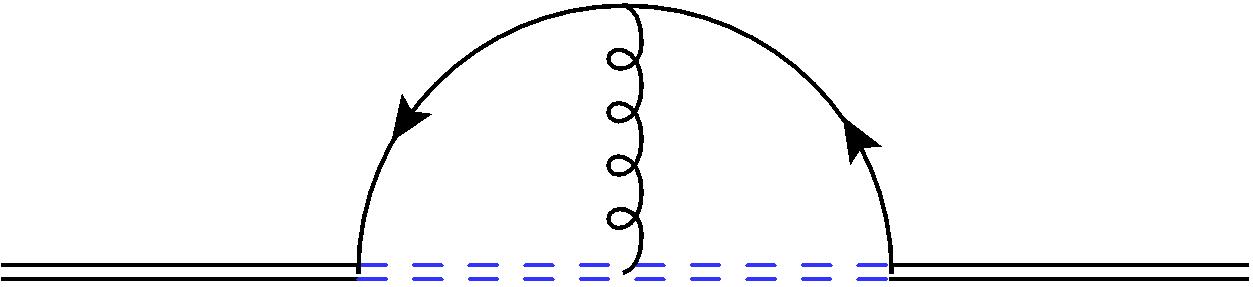}}} &=  -\bar\alpha\alpha \frac{g^4N^2}{16\pi^2} (c_z)^{2\epsilon} \, \frac{1}{\epsilon} \, \left[ \frac{1}{2\epsilon} + \log\left( 4\pi e^{\gamma_E-2} \right)  \right] \int  ds \ \bar z \dot z  \,,\nonumber
\end{split}
\end{equation}
plus the one-loop counterterms
\begin{equation}
\begin{split}
    \vcenter{\hbox{\includegraphics[width=0.15\textwidth]{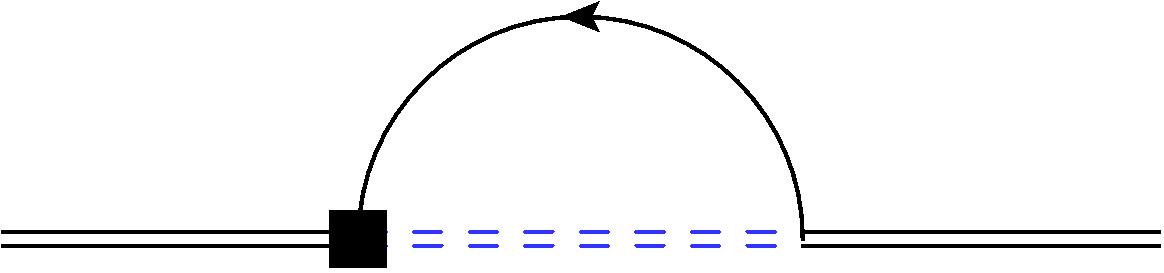}}} &= \bar\alpha\alpha \frac{g^4N^2}{16\pi^2} (c_z)^{2\epsilon} \, \frac{1}{\epsilon} \, \left[ \frac{1}{\epsilon} + \log\left( 4\pi e^{\gamma_E-2}\right) \right]\int ds \ \bar z\dot z \,,\\
    \vcenter{\hbox{\includegraphics[width=0.15\textwidth]{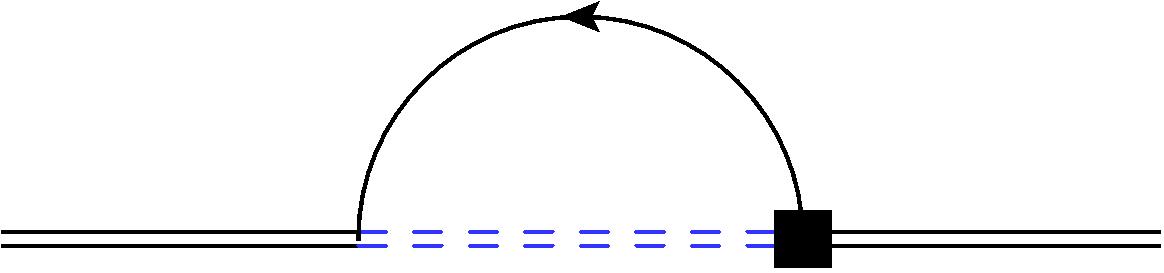}}} &= \bar\alpha\alpha \frac{g^4N^2}{16\pi^2}(c_z)^{2\epsilon} \, \frac{1}{\epsilon} \, \left[ \frac{1}{\epsilon} + \log\left( 4\pi e^{\gamma_E-2}\right) \right]\int ds \ \bar z\dot z \,,\\
    \vcenter{\hbox{\includegraphics[width=0.15\textwidth]{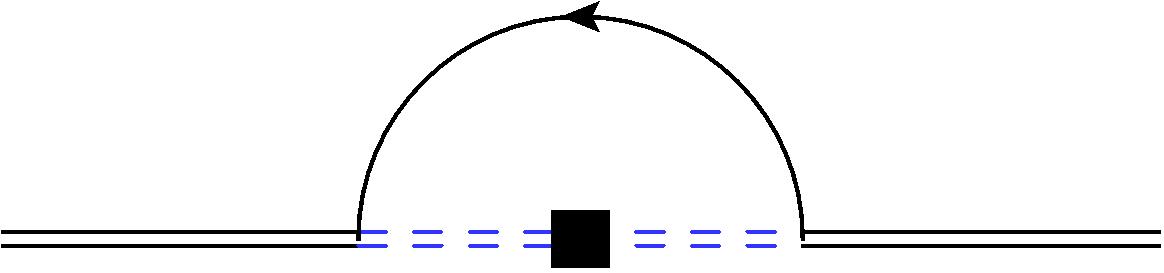}}} &= -\bar\alpha^2\alpha^2 \frac{g^4N^2}{16\pi^2}(c_z)^{2\epsilon} \, \frac{1}{\epsilon} \, \left[ \frac{1}{\epsilon} + \log\left( 4\pi e^{\gamma_E-2} \right) \right]\int ds \ \bar z\dot z  \,.
\end{split}
\end{equation}
Imposing the counterterms to cancel the divergences (up to scheme dependent finite terms) we find 
\begin{equation}
\label{eq:Zz}
\begin{split}
    Z_z \equiv 1 + \delta_z = 1 - \bar\alpha\alpha \frac{g^2N}{4\pi} \, \frac{(c_z)^{\epsilon}}{\epsilon} -\bar\alpha\alpha \frac{g^4N^2}{16\pi^2}\, \frac{(c_z)^{2\epsilon}}{\epsilon} \,  \left[ \frac{(\bar\alpha\alpha-2)}{2\epsilon} + (\bar\alpha\alpha-1) \right] \,.
\end{split}
\end{equation}
The same result holds for all the other one-dimensional fields  and coincides with the renormalization function $Z_\Psi$ of the $\Psi$ supermatrix in \eqref{eq:oddmatrix}.

\subsection{Renormalization of the parameters}

At one loop, bare and renormalized parameters are related as \cite{Castiglioni:2022yes}
\begin{equation}
\label{eqn:renormalizedalphas}
\begin{split}
    \alpha_0 &= \left[ 1+g^2\frac{N}{4\pi} \, (\bar\alpha \alpha-1) \,\frac{(c_z)^\epsilon}{\epsilon} \, \right]\alpha\,, \\ \bar\alpha_0 &= \left[ 1+g^2\frac{N}{4\pi} \, (\bar\alpha \alpha-1) \,\frac{(c_z)^\epsilon}{\epsilon} \,  \right]\bar\alpha\, .
\end{split}
\end{equation}
where we have included the scheme factor $c_z$. 
Reading the definition of the renormalization function $Z_\alpha$ in \eqref{eq:parren} and substituting the result for $Z_z$ in \eqref{eq:Zz} up to one loop, we obtain
\begin{equation}\label{eq:Za}
        Z_{\bar\alpha} = Z_\alpha \equiv 1 + \delta_\alpha = 1  -g^2 \frac{N}{4\pi} \,\frac{(c_z)^{\epsilon}}{\epsilon}\,.
\end{equation}

From the renormalization of the $\bar\alpha \alpha$ parameter, 
\begin{equation}
   \bar\alpha \alpha = Z_\alpha^{-2} Z_z^2 \, \bar\alpha_0 \alpha_0 
\end{equation}
one can easily obtain the following $\beta$-function
\begin{equation}\label{eq:beta}
\begin{split}
    \beta_{\bar\alpha\alpha} &=  \bar\alpha\alpha(\bar\alpha\alpha-1)\frac{g^2N}{\pi}(c_z)^{\epsilon}\,,\\
    \beta_{g^2} &= -2\epsilon g^2 \, ,
\end{split}
\end{equation}
where, for later convenience, we have included also the trivial $\beta_{g^2}$ of the ABJM coupling. 
We stress that these expressions are valid for any (open or closed) contour, since renormalization, being performed locally, cannot be affected by 
the shape of the path.

The appearance of non-vanishing $\beta$-functions induces non-trivial RG flows, which connect two fixed points, 
the $\bar\alpha \alpha =0$ one corresponding to the cusped version of the 1/6 BPS bosonic line and $\bar\alpha \alpha =1$ describing the cusped version of th 1/2 BPS line.

Solving equation \eqref{eq:beta} for the running $\bar\alpha\alpha$ coupling we obtain
\begin{equation}\label{eq:beta_solution}
    \bar\alpha\alpha(\mu) = \frac{1}{1 + \left(\mu/\Lambda\right)^{\frac{g^2N}{\pi}} } \,,
\end{equation}
where $\Lambda$ is a boundary integration scale.

\section{The interpolating generalized cusp}
\label{sec:Cusp}

We now introduce a {\em parametric} cusped line operator. Its construction relies on the generalization of the original setting \textcolor{red}{\cite{Griguolo:2012iq}}, where we incorporate a parametric dependence through the ``hyperloop prescription'' \cite{Drukker:2019bev,Drukker:2020dvr}.

To this end, we first review the general construction of parametric 1/6 BPS fermionic operators given in \cite{Drukker:2019bev,Drukker:2020dvr}. The starting point is the bosonic operator defined along the line
\begin{equation}
\label{eq:lineWL0}
    W=\frac{1}{2N} \, \Tr\cP \bigg(e^{-i\int \cL_0 \,ds}\bigg)\, ,
\end{equation}
with connection given by
\begin{equation}
\label{eq:bosonicA}
    \cL_0=\begin{pmatrix}
        \cA & 0 \\
        0 & \hat \cA 
    \end{pmatrix}\, , \qquad \begin{cases} \cA \equiv A_{\mu}\dot x^{\mu}-ig^2 |\dot x|M_J^{\ I}C_I \bar C^J\\
    \hat\cA \equiv A_{\mu}\dot x^{\mu}-ig^2 |\dot x|M_J^{\ I}\bar C^J C_I\\
\end{cases} \quad (g^2 \equiv 2\pi/k) \, ,
\end{equation}
with $M=\diag(-1,-1,1,1)$. Here $N$ is the rank of the gauge groups of the two nodes of the ABJM quiver and $k$ is the Chern-Simons level (for conventions on ABJM see appendix \ref{app:abjm}). This operator is annihilated by two Poincar\'e supercharges ($Q_{12}^+$ and $Q_{34}^-$)
as well as two superconformal ones ($S_{12}^+$ and $S_{34}^-$). It is therefore $1/6$ BPS. 

The fermionic line, preserving the same supercharges, is obtained by taking their sum ${\cal Q}= Q_{12}^+ + Q_{34}^-$  and defining the following superconnection 
\begin{equation}
\label{eq:superconnection}
    \cL = \cL_0+i\cQ G + G^2\,,
\quad {\rm with} \qquad
    G =\sqrt{2i}g \begin{pmatrix}
        0 && \bar\alpha \, C_2 \\
        \alpha \, \bar{C}^2 && 0
    \end{pmatrix}.
\end{equation}

The resulting operator is parameterized by two complex (but not complex conjugates) parameters $\bar\alpha$ and $\alpha$, and it is explicitly written as
\begin{equation}
\label{eq:lineWL}
    W= \frac{1}{2N} \Tr\cP \bigg(e^{-i\int \cL \,ds}\bigg)\,,\quad \cL = \begin{pmatrix}
        \cA && \bar{f}\\
        f && \hat{\cA}
    \end{pmatrix}\,,
\end{equation}
with $\cA$ and $\hat\cA$ defined as \eqref{eq:bosonicA}, now with $M = \diag(-1,-1+2\bar\alpha \alpha,1,1)$
and off-diagonal elements $\bar{f} = -g\sqrt{2i}\, \bar\alpha\bar\psi^1_+$ and $f = g\sqrt{2i}\, \alpha \psi_1^+$. Details on the construction can be found in \cite{Drukker:2019bev,Drukker:2020dvr}.

For generic values of $\bar\alpha$ and $\alpha$, the fermionic operator preserves all supercharges preserved by the $1/6$ bosonic loop \eqref{eq:lineWL0} \cite{Ouyang:2015iza}. It then describes a family of fermionic 1/6 BPS line operators. At the particular point $\bar\alpha\alpha=1$, R-symmetry is enhanced from $SU(2)$ to $SU(3)$ and we gain 8 extra preserved supercharges (4 Poincar\'e + 4 superconformal). Together with the original 2+2 supercharges, they lead to a $1/2$ BPS line operator. 

\begin{figure}[H]
    \centering    \includegraphics[width=\textwidth]{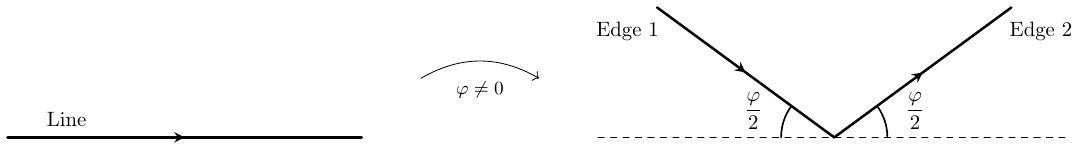}
    \caption{Deforming the line to a cusp with  angle $\varphi$.}
    \label{fig:linetocusp}
\end{figure}

We now consider the cusped contour 
\begin{equation}
\label{eq:cuspedcontour}
    x^{\mu} = \left(0,s \cos\tfrac{\varphi}{2}, |s| \sin\tfrac{\varphi}{2}  \right) \,,
\end{equation}
where $\varphi$ is a cusp angle, as shown on the right side of figure \ref{fig:linetocusp}, and $s$ is a real parameter. Edge 1 is parameterized by $s<0$, while edge 2 corresponds to $s>0$. 

In the original setting of \cite{Griguolo:2012iq}, the name ``generalized cusp" refers to the Wilson operator \eqref{eq:lineWL} with $\bar\alpha \alpha =1$, supported on path \eqref{eq:cuspedcontour},  also characterized by a relative R-symmetry rotation in the $M_J^{\;\; I}$ matrices in \eqref{eq:bosonicA} that encodes the couplings to the matter fields on the two edges\footnote{In the dual description this corresponds to a fundamental string with a ``jump" in ${\mathbb C}{\text P}^3$ \cite{Forini:2012bb}.}. The resulting operator depends on the cusp angle $\varphi$ and the internal R-rotation angle $\theta$.

If we now generalize this operator to the case of generic $\bar{\alpha},\alpha$ 
we obtain a generalized cusp of the form \eqref{eq:lineWL}, supported on path \eqref{eq:cuspedcontour} and featured by two different superconnections ${\cal L}_1$ and ${\cal L}_2$ on edge 1 and 2, respectively, which depend on $\bar\alpha, \alpha$ in addition to the $\varphi, \theta$ angles. Precisely, the bosonic entries  \eqref{eq:bosonicA} depend on the following scalar coupling matrices\footnote{In principle, one could introduce different parameters $\bar\alpha_1,\alpha_1$ and $\bar\alpha_2,\alpha_2$ on the two edges. However, we restrict to the same set of parameters, as we want to recover the parametric 1/6 BPS line operator in the $\varphi, \theta \to 0$ limit.} 

\begin{equation}
\label{eq:Mscusp}
\begin{split}
     &(M_1)_I^{\ J} = \begin{pmatrix}
        -\cos\tfrac{\theta}{2} && 0 && -\sin\tfrac{\theta}{2} && 0 \\
        0 && -1+2\bar\alpha\alpha && 0 && 0 \\
       -\sin\tfrac{\theta}{2} && 0 && \cos\tfrac{\theta}{2} && 0\\
        0 && 0 && 0 && 1
    \end{pmatrix} \,,\\ & (M_2)_I^{\ J} = \begin{pmatrix}
        -\cos\tfrac{\theta}{2} && 0 && \sin\tfrac{\theta}{2} && 0 \\
        0 && -1+2\bar\alpha\alpha && 0 && 0 \\
       \sin\tfrac{\theta}{2} && 0 && \cos\tfrac{\theta}{2} && 0\\
        0 && 0 && 0 && 1
    \end{pmatrix} \,,
\end{split}
\end{equation}
while the fermionic entries are given by\footnote{Contraction of spinorial indices is always taken to be up-down. For example $\eta \bar\psi \equiv \eta^\delta \bar\psi_\delta$.}
\begin{equation}
\label{eq:fscusp}
 \bar{f}_a = -g \sqrt{i} \bar\alpha (\eta_a)_I \bar\psi^I\,, \qquad f_a = g \sqrt{i} \alpha \psi_I (\bar\eta_a)^I\,, \quad a=1,2\,,
\end{equation}
where
\begin{equation}
\begin{split}
   & (\eta_1)^{\delta}_I = (e^{-i\frac{\varphi}{4}}, e^{i\frac{\varphi}{4}})^{\delta} \left( \cos\tfrac{\theta}{4}, 0,\sin\tfrac{\theta}{4},0 \right)_I \,, \qquad\quad (\bar\eta_1)_{\delta}^I = \begin{pmatrix}
        e^{i\frac{\varphi}{4}} \\ e^{-i\frac{\varphi}{4}}
    \end{pmatrix}_{\!\!\! \delta}\begin{pmatrix} \cos\tfrac{\theta}{4} \\ 0 \\ \sin\tfrac{\theta}{4} \\0
    \end{pmatrix}^{\!\!\!\!\! I}\,, \\ &
    (\eta_2)^{\delta}_I = (e^{i\frac{\varphi}{4}}, e^{-i\frac{\varphi}{4}})^{\delta} \left( \cos\tfrac{\theta}{4} ,0,- \sin\tfrac{\theta}{4},0 \right)_I \,, \qquad (\bar\eta_2)_{\delta}^I = \begin{pmatrix}
        e^{-i\frac{\varphi}{4}} \\ e^{i\frac{\varphi}{4}}
    \end{pmatrix}_{\!\!\! \delta}\begin{pmatrix} \cos\tfrac{\theta}{4} \\ 0 \\ -\sin\tfrac{\theta}{4} \\0      
    \end{pmatrix}^{\!\!\!\!\! I}\,.
\end{split}
\end{equation}

This is a new family  of parametric cusped operators, corresponding to the cusped version of $1/6$ BPS fermionic representatives, that we call ``interpolating generalized cusp". In fact, varying $\bar\alpha, \alpha$ it interpolates between the cusped version of the $1/6$ BPS bosonic (at $\bar\alpha =\alpha =0$) and the cusped version of  the $1/2$ BPS line (at $\bar\alpha\alpha=1$). When supported along the cusped path \eqref{eq:cuspedcontour}, supersymmetry is generally broken and the operators are no longer BPS. Nevertheless, for simplicity, we will refer to them with the fraction of supersymmetry preserved by the corresponding Wilson lines in the zero-cusp limit.


\subsection{Cusped Wilson line via one-dimensional theory}
\label{sec:cuspedWL}

The presence of a cusp on the Wilson line contour gives rise to short distance singularities, and the corresponding VEV needs to be appropriately renormalized. As a consequence, the operator acquires an anomalous dimension $\Gamma_{\rm cusp}$, usually called \emph{cusp anomalous dimension}. When exponentiation theorems \cite{Dotsenko:1979wb,Gatheral:1983cz,Frenkel:1984pz} are at work, $\Gamma_{\rm cusp}$ is
given by the coefficient of the exponentiated divergent term,
\begin{equation}\label{eq:cuspandim}
    \langle W_{\rm cusp} \rangle = e^{-\Gamma_{\rm cusp}(\theta,\varphi)\log \frac{\Lambda}{\mu}} + \text{finite terms}\,,
\end{equation}
where $\Lambda$ is an IR cutoff and $\mu$ stands for the renormalization scale. In dimensional regularization ($d \to d -2\epsilon$) UV divergences appear in the exponent as simple poles in $\epsilon$, with  $\Gamma_{\rm cusp}$ being the corresponding residue. 

$\Gamma_{\text{cusp}}$ can be perturbatively determined from the renormalized $\langle W_{\rm cusp} \rangle$, defined as
\begin{equation}\label{eq:WcuspR1}
    \langle W_{\text{cusp}} \rangle = Z_{\text{open}}^{-1} Z_{\text{cusp}}^{-1} \langle W_{\rm cusp} \rangle_0 \,,
\end{equation}
where $\langle W_{\rm cusp} \rangle_0  $ is the bare (divergent) VEV, $Z_{\text{open}}$ is the renormalization function that ensures that in the $\theta,\varphi \to 0$ limit the normalized VEV of the straight Wilson line is recovered ($\langle W_{\text{line}}\rangle=1$), and $Z_{\text{cusp}}$ is the cusp renormalization function which should cure the remaining UV divergences. According to the standard prescription, the cusp anomalous dimension is then given by
\begin{equation}\label{eq:gammacusp}
    \Gamma_{\text{cusp}} = \mu \frac{d}{d \mu} \log Z_{\text{cusp}}\,.
\end{equation}

In order to evaluate $\Gamma_{\text{cusp}}$ for the parametric cusped Wilson line introduced in the previous section, we will first evaluate $Z_{\rm cusp}$ by 
generalizing the one-dimensional auxiliary theory reviewed in section \ref{app:1dmethod} to include cusped contours. As we will see, in the presence of a cusp the one-dimensional action has to be adapted to incorporate the path singularity at the cusp. 

We consider the path in figure \ref{fig:linetocusp}, with the cusp located at the origin, $s = 0$. In order to tame IR divergences, the two half lines are cut to finite length $L$, the left one being parametrized by $-L < s <0$ and the right one by $0 < s < L$. In principle, the IR cut-off has to be removed at the end of the calculation, sending $L$ to infinity. However, while this would be a safe operation once the perturbative series has been resummed \cite{Chishtie:2017vwk}, at any finite order in perturbation theory $L$-dependent terms  can be present, whenever conformal invariance is broken. We will carefully address this question in our perturbative results.

The action for the one-dimensional auxiliary theory reads 
\begin{equation}
\label{eq:effactionwithsource}
    S_{\text{eff}} = S_{\text{ABJM}} + \int_{-L}^L ds \Tr \left[\bar\Psi\left( \partial_{s} + i\cL_1\Theta(-s) + i \cL_2\Theta(s) \right)\Psi + \lambda\bar\Psi\Psi \delta(s+L) \right]\,.
\end{equation}
where $S_\text{ABJM}$ is the the action of the bulk ABJM theory (see \eqref{eq:ABJMaction}) and $\Psi$ is the one-dimensional auxiliary supermatrix defined in \eqref{eq:oddmatrix}. 
For $\lambda = 1$ the operator localized in $s=-L$ makes the action invariant under charge conjugation plus inversion of the path ordering. It is a manifestation of the presence of the IR regulator $L$. The inclusion of the $\lambda$ coupling takes into account possible quantum corrections to this composite operator. 

In the presence of the IR cutoff $L$, definition \eqref{eq:gammacusp} has to be modified as
\begin{equation}\label{eq:gammacuspL}
    \Gamma_{\text{cusp}} = \left(\mu\frac{d}{d\mu} - L \frac{d}{d L}\right)\log Z_{\text{cusp}}\,.
\end{equation}
As we will discuss later, the $L$-derivative is necessary to  remove from $\Gamma_{\text{cusp}}$ unwanted boundary effects.

According to the general prescription of \cite{Dorn:1986dt, Castiglioni:2022yes}, for a {\em smooth} contour the VEV of the bare Wilson line operator equals the two-point function $\Tr\langle \Psi_0(L)\bar\Psi_0(-L)\rangle$, where $\Psi_0, \bar\Psi_0$ are the bare one-dimensional fields. It follows that if the regular contour is split in $s_0$ into two segments, we can write 
\begin{equation}\label{eq:split1}
    \langle \Psi_0(L)(\bar\Psi_0\Psi_0)(s_0)\bar\Psi_0(-L)\rangle = \langle \Psi_0(L)\bar\Psi_0(s_0) \rangle \langle \Psi(s_0)\bar\Psi_0(-L)\rangle=\langle \Psi_0(L)\bar\Psi_0(-L)\rangle\,.
\end{equation}
Renormalizing both sides of \eqref{eq:split1}, one can prove that the composite operator  $\bar\Psi_0\Psi_0$ in $s_0$ does not renormalize. 

Instead, if a {\em cusp} is present at $s_0$ ($s_0=0$ in our case), a non-trivial renormalization of the composite operator localized at the cusp arises, due to the appearance of short distance singularities close to the cusp. Therefore, for the renormalized VEV we write
\begin{equation}\label{eq:WcuspR2}
    \langle W_{\text{cusp}}\rangle = \frac{1}{2}\Tr\langle \Psi(L) \left[ \bar\Psi \Psi \right](0) \bar\Psi(-L) \rangle = \frac{1}{2} Z_{\Psi}^{-1} Z_{\bar\Psi\Psi}^{-1} \Tr\langle \Psi_0(L) ( \bar \Psi_0  \Psi_0)(0) \bar\Psi_0(-L) \rangle\,,
\end{equation}
where we have defined the renormalized parameters as
\begin{equation}\label{eq:ren}
    \Psi = Z_\Psi^{-1/2} \Psi_0 \, , \qquad 
    \bar\Psi = Z_\Psi^{-1/2} \bar\Psi_0 \, , \qquad
    \left[\bar\Psi \Psi \right]= Z_{\bar \Psi \Psi}^{-1} \bar\Psi_0 \Psi_0 = Z_{\bar \Psi \Psi}^{-1} Z_\Psi \, \bar\Psi \Psi \, .
\end{equation}
Matching \eqref{eq:WcuspR2} with the usual renormalization \eqref{eq:WcuspR1}  of a cusped Wilson loop leads to identifying the renormalization functions of the one-dimensional field and the composite operator with $Z_{\text{open}}$ and $Z_{\text{cusp}}$, respectively\footnote{In the ABJM theory we have a single $Z_{\rm cusp}$ and consequently a single cusp anomalous dimension. In the more general $U(N_1) \times U(N_2)$ ABJ case, one should define two cusp anomalous dimensions, which differ simply by the exchange of $N_1$ with $N_2$ \cite{Griguolo:2012iq}.} 
\begin{equation}\label{eq:Zdef}
    Z_{\text{open}} = Z_{\Psi} \,, \qquad \qquad Z_{\text{cusp}} = Z_{\bar\Psi\Psi}\,.
\end{equation}
Therefore, in the one-dimensional theory formalism the cusp renormalization function corresponds to the renormalization function of the localized composite operator $\bar\Psi\Psi (0)$. In what follows we focus on the perturbative evaluation of $Z_{\bar\Psi\Psi}$. 

Along the calculation we will make use of the two-loop result for $Z_\Psi$ ($\equiv Z_z$) given in \eqref{eq:Zz}. As previously mentioned, this  expression is scheme dependent, with scheme dependence being encoded in an arbitrary constant $c_z$.


\subsection{Perturbative renormalization }

To compute the renormalization function of $\bar\Psi \Psi (0)$, we source the composite operator by a supermatrix  $J_0$ adding to the bare action \eqref{eq:effactionwithsource} the term
\begin{equation}
    \int_{-L}^L ds \ \Tr \left( J_0 \bar\Psi_0 \Psi_0 \right) \, \delta(s)\,.
\end{equation}
Using ordinary BPHZ renormalization we 
rewrite it as
\begin{equation}\label{eq:Jvertex}
    \int_{-L}^L ds \ \Tr \left( J [\bar\Psi \Psi]\right) \,  \delta(s) + \text{counterterms}\,,
\end{equation}
where we have defined $J = Z_J^{-1} J_0$ and expressed  everything in terms of renormalized quantities. From the identity $J_0 \bar\Psi_0 \Psi_0 = J [\bar\Psi \Psi]$,
it follows that $Z_{\bar\Psi \Psi} = Z_{J} ^{-1}$.
We then trade the evaluation of $Z_{\bar\Psi \Psi}$ with the evaluation of $Z_J$, which in turn can be read from the renormalization of the vertex in \eqref{eq:Jvertex}. 

We can simplify the calculation by focusing only on the first entry of the $J$ supermatrix (still called $J$), that is on the $J \bar z z$ vertex. 
We write
\begin{equation}
    J_0 \bar z_0 z_0 = Z_{J} Z_{z} J \bar z z \equiv Z_{v} \, J \bar z z = (1 + \delta_v) \, J \bar z z \,.
\end{equation}
From the previous definitions it follows that 
\begin{equation}
    Z_{\text{cusp}} = Z_{\bar z z} = Z_{z}/Z_v \,.
\end{equation}
In perturbation theory we determine $Z_z$ and $Z_{v}$ separately. The renormalization function $Z_z$ of the elementary field, up to two loops, is given in \eqref{eq:Zz}. Here instead we focus on the evaluation of the cusp renormalization function $Z_{v}$. Conventions for Feynman diagrams for the one-dimensional theory are collected in section \ref{app:1dmethod}.

\subsection*{One-loop diagrams}

At one loop there is only one divergent correction to the $J\bar zz$ vertex, coming from a diagram with a fermion exchange between two different edges. The evaluation of the corresponding integral can be found in \cite{Griguolo:2012iq} (see their equation (4.12)). Exploiting that result, while including the $\alpha,\bar\alpha$ dependence and a suitable (but totally generic) scheme factor $c_J$, we find 
\begin{equation}
\label{eq:oneloopintegral}
    \vcenter{\hbox{\includegraphics[width=0.11\textwidth]{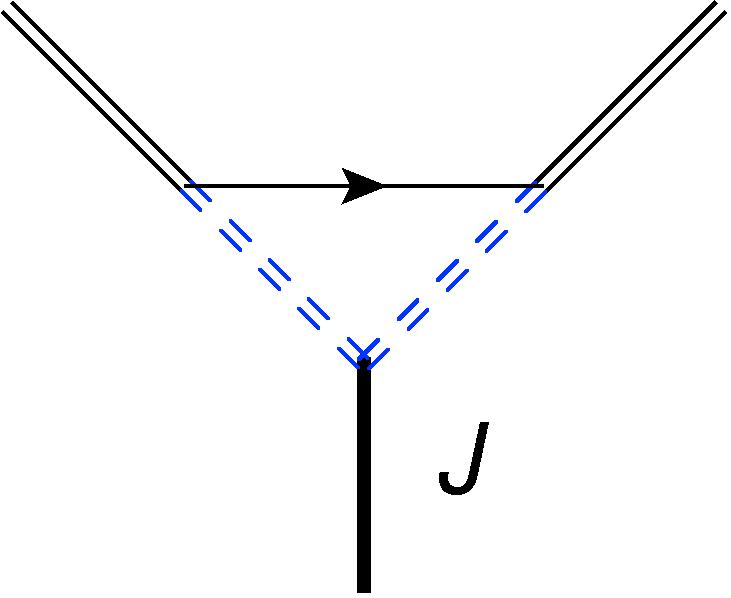}}} = -\bar\alpha\alpha \frac{g^2N}{4\pi}  \frac{\cos\frac{\theta}{2}}{\cos\frac{\varphi}{2}} \, \frac{1}{{\epsilon}} \, (c_J)^{\epsilon}(\mu L)^{2\epsilon} \, J \bar z z(0)\,.
\end{equation}
The one-loop counterterm 
\begin{equation}
    \vcenter{\hbox{\includegraphics[width=0.11\textwidth]{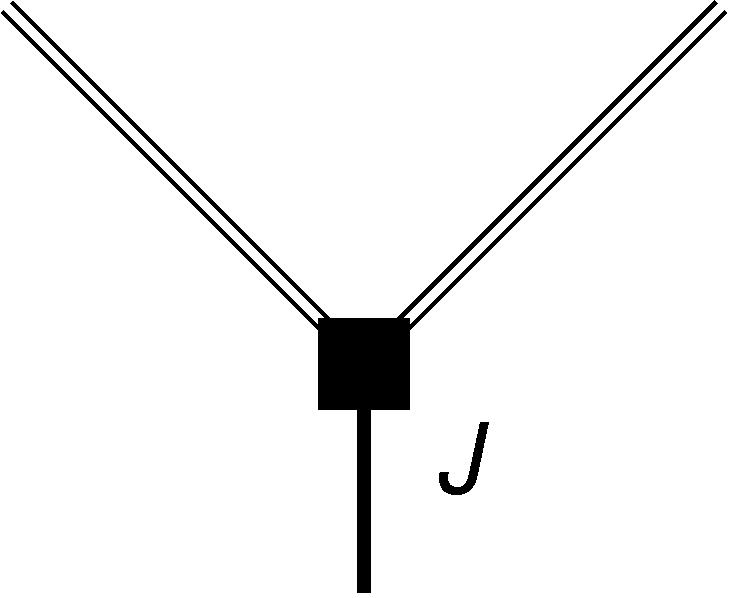}}} = -\delta_v J \bar z z(0)
\end{equation}
needed to cancel this divergence, leads eventually to the following one-loop renormalization function
\begin{equation}\label{eq:deltav}
    Z_v = 1+ \delta_v = 1 -\bar\alpha\alpha \frac{g^2N}{4\pi}  \frac{\cos\frac{\theta}{2}}{\cos\frac{\varphi}{2}}\frac{1}{{\epsilon}} \, (c_J \mu^2 L^2)^\epsilon\,.
\end{equation}

\subsection*{Two-loop diagrams}

We split two-loop divergent diagrams into purely bosonic, fermionic and counterterm insertions. A detailed evaluation of the corresponding integrals can be found in \cite{Griguolo:2012iq}. We import those results by including the $\alpha,\bar\alpha$ dependence and a $c_J$ scheme factor associated to each loop integral, as well. 

From the first class of diagrams we have two contributions that sum up to
\begin{equation}
    \vcenter{\hbox{\includegraphics[width=0.11\textwidth]{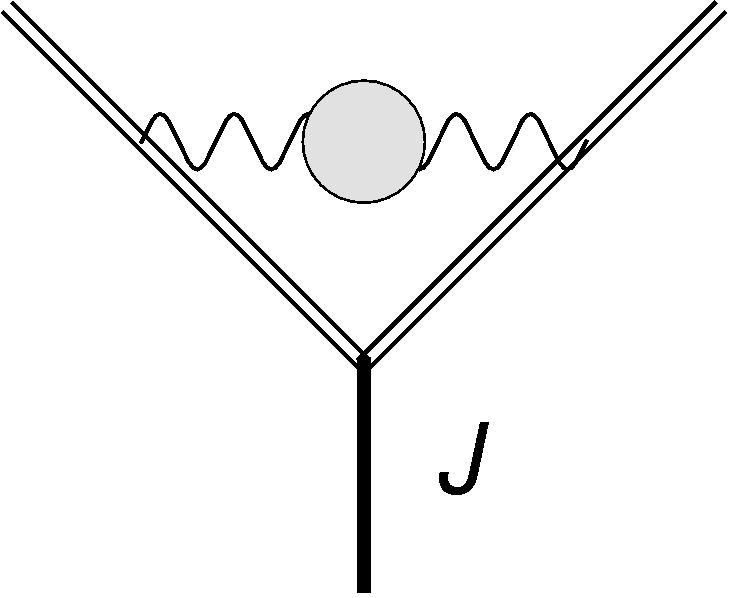}}} + \vcenter{\hbox{\includegraphics[width=0.11\textwidth]{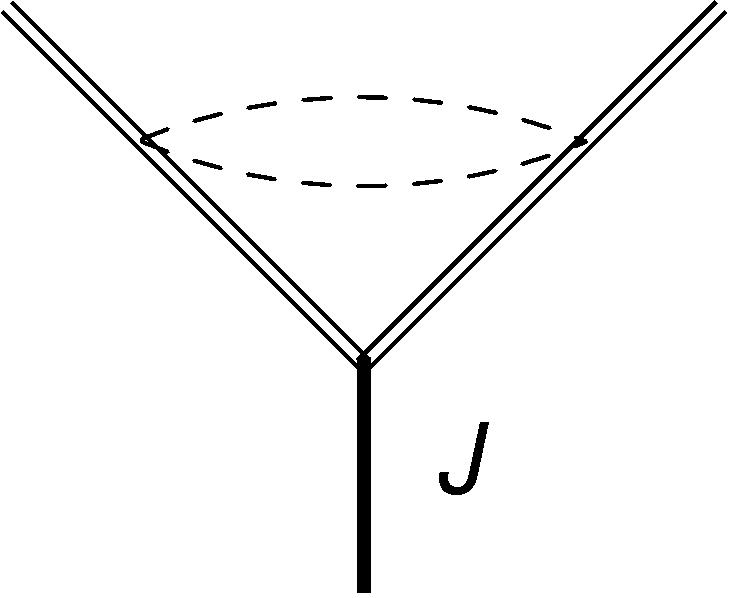}}} = -\frac{g^4N^2}{16\pi^2} \, \frac{1}{\epsilon} \,  (c_J)^{2\epsilon}(\mu L)^{4\epsilon}\frac{\varphi}{\sin\varphi} \left( \cos^2\tfrac{\theta}{2} -\cos\varphi +  \alpha\bar\alpha(\alpha\bar\alpha-1) \right) J\bar z z(0)\,.
\end{equation}

From diagrams with fermionic corrections we obtain
\begin{equation}
\begin{split}
    \vcenter{\hbox{\includegraphics[width=0.11\textwidth]{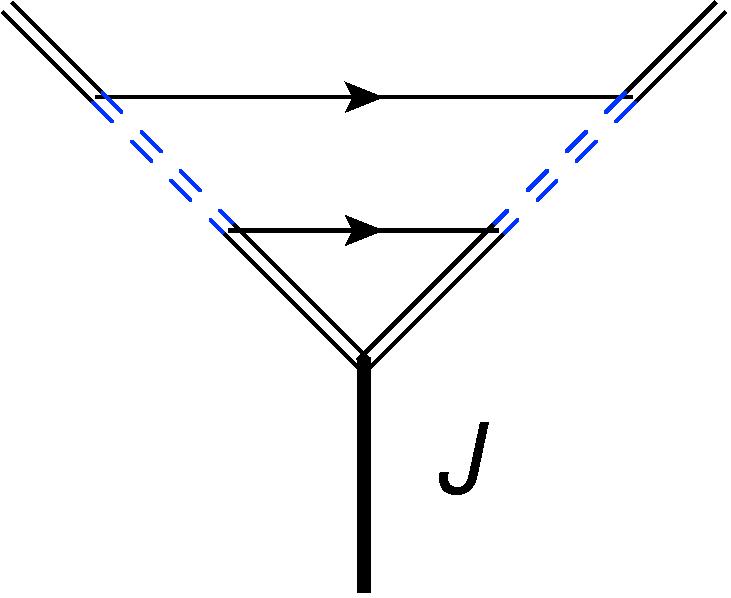}}} &= -\bar\alpha\alpha\frac{g^4N^2}{16\pi^2}\, \frac{1}{\epsilon} \,  (c_J)^{2\epsilon}(\mu L)^{4\epsilon}\frac{\cos\frac{\theta}{2}}{\cos\frac{\varphi}{2}} \Bigg[ 
    \frac{1}{2\epsilon}  +\log\left( 4\pi e^{\gamma_E} \right) \\ & \qquad\qquad\qquad\qquad\qquad\qquad-2\log\left(1+\sec\frac{\varphi}{2}\right)  -\frac{\varphi}{2}\cot\frac{\varphi}{2} \Bigg] J\bar z z(0) \,,\\
    \vcenter{\hbox{\includegraphics[width=0.11\textwidth]{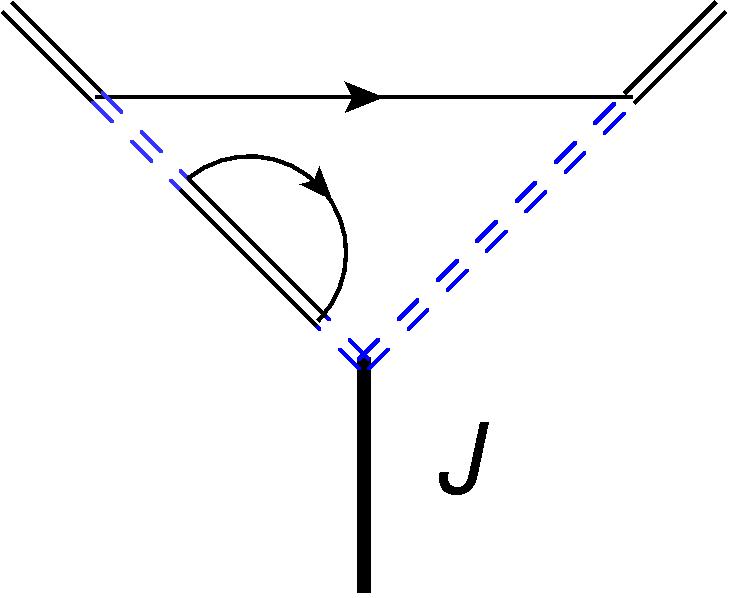}}} &=\vcenter{\hbox{\includegraphics[width=0.11\textwidth]{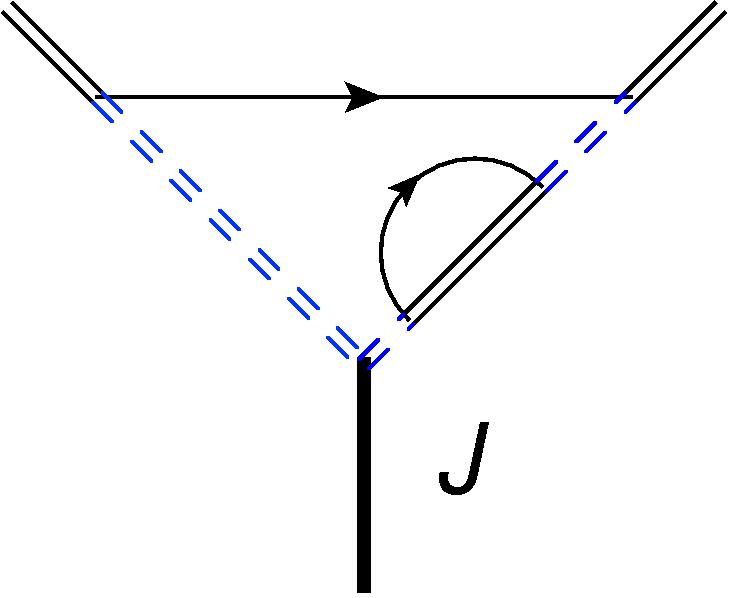}}}= \bar\alpha^2\alpha^2\frac{g^4N^2}{16\pi^2}\, \frac{1}{\epsilon} \, (c_J)^{2\epsilon}(\mu L)^{4\epsilon}\frac{\cos\frac{\theta}{2}}{\cos\frac{\varphi}{2}}\Bigg[ \frac{1}{2\epsilon} + \log\left( 4\pi e^{\gamma_E} \right)  \\ & \qquad\qquad\qquad\qquad\qquad\qquad\qquad+ \log\left( \frac{1}{4}\cos\frac{\varphi}{2}\sec^4\frac{\varphi}{4} \right) \Bigg] J\bar z z(0) \,,\\
    \vcenter{\hbox{\includegraphics[width=0.11\textwidth]{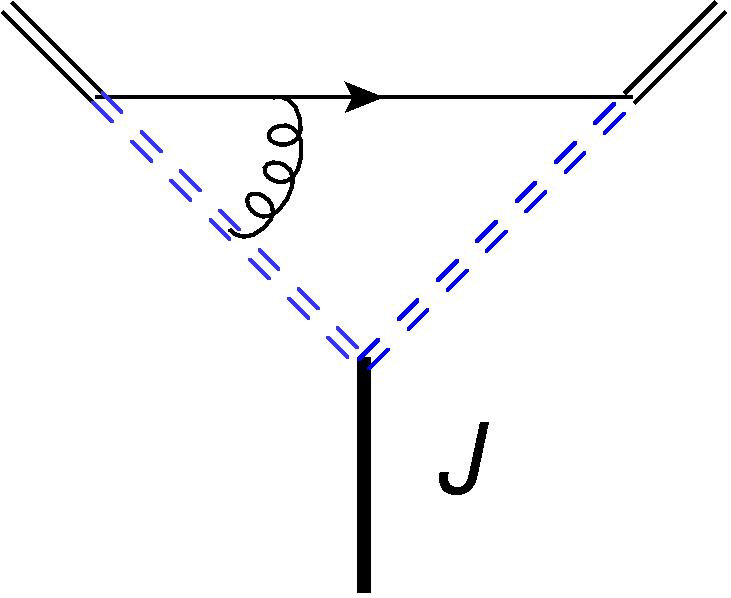}}} &=\vcenter{\hbox{\includegraphics[width=0.11\textwidth]{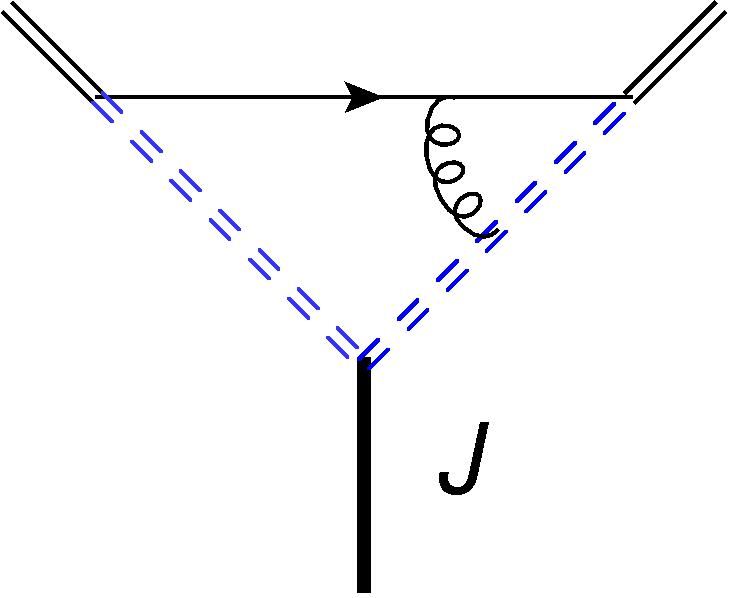}}}= -\bar\alpha\alpha\frac{g^4N^2}{32\pi^2} \, \frac{1}{\epsilon} \, (c_J)^{2\epsilon}(\mu L)^{4\epsilon}\frac{\cos\frac{\theta}{2}}{\cos\frac{\varphi}{2}} \Bigg[ 
    \frac{1}{2\epsilon}  +\log\left( 4\pi e^{\gamma_E} \right)  \\ & \qquad\qquad\qquad\qquad\qquad\qquad-2\log\left(1+\sec\frac{\varphi}{2}\right)  +\frac{\varphi}{4}\cot\frac{\varphi}{2} \Bigg] J\bar z z(0) \,,\\
    \vcenter{\hbox{\includegraphics[width=0.11\textwidth]{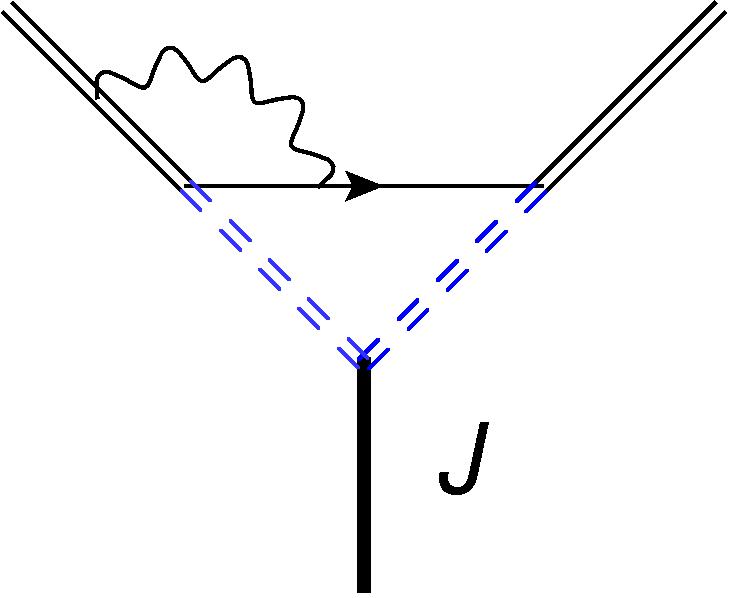}}} &=\vcenter{\hbox{\includegraphics[width=0.11\textwidth]{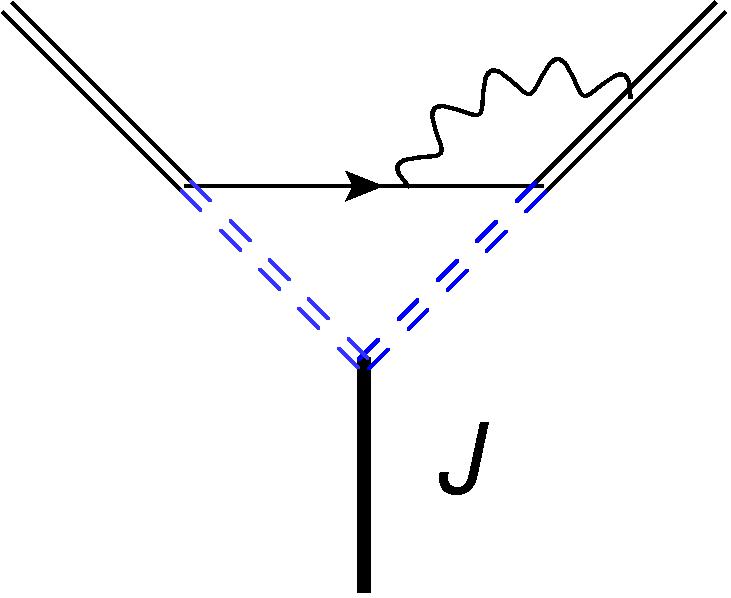}}}= -\bar\alpha\alpha \frac{g^4 N^2}{32\pi^2}\, \frac{1}{\epsilon} \, (c_J)^{2\epsilon}(\mu L)^{4\epsilon}\frac{\cos\frac{\theta}{2}}{\cos\frac{\varphi}{2}} \Bigg[ 
    \frac{1}{2\epsilon} +\log\left( 4\pi e^{\gamma_E} \right) \\ & \qquad\qquad\qquad\qquad\qquad\qquad-2\log\left(1+\sec\frac{\varphi}{2}\right)  -\frac{\varphi}{4}\cot\frac{\varphi}{2} \Bigg]J \bar z z(0) \,,\\
    \vcenter{\hbox{\includegraphics[width=0.11\textwidth]{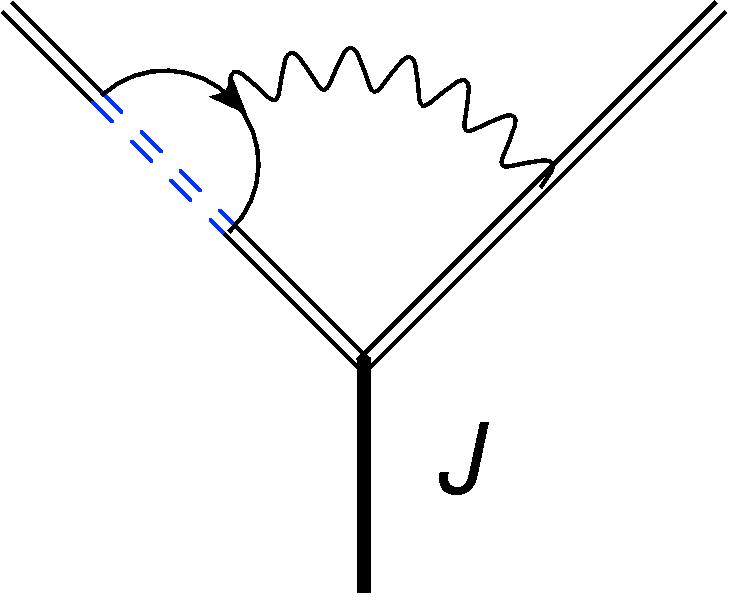}}} &=\vcenter{\hbox{\includegraphics[width=0.11\textwidth]{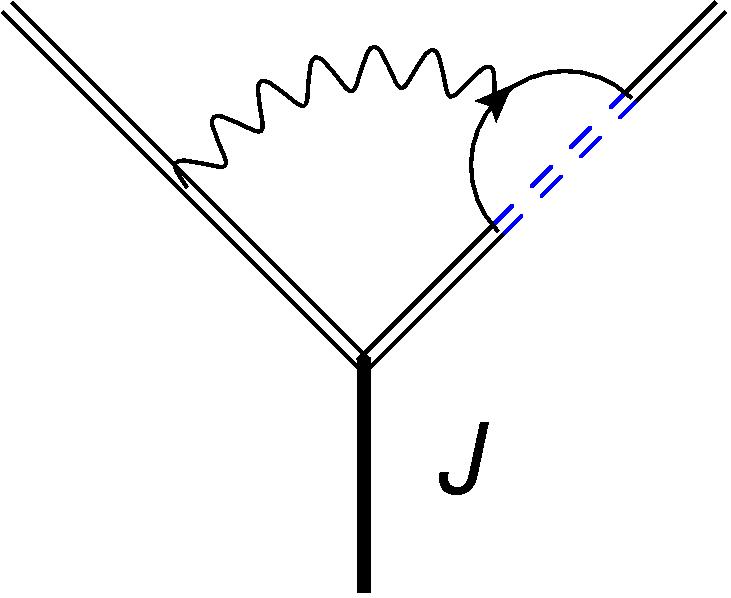}}}= \bar\alpha\alpha\frac{g^4N^2}{16\pi^2}\, \frac{1}{\epsilon} \, (c_J)^{2\epsilon}(\mu L)^{4\epsilon}\left[ \log\left( \cos\frac{\varphi}{2} \right) -\frac{\varphi}{2}\cot\varphi \right] J\bar z z(0) \,,
\end{split}
\end{equation}

In addition, there are contributions due to the insertion of one-loop counterterms $\delta_z$, $\delta_\alpha$, eqs. (\ref{eq:Zz}, \ref{eq:Za}),
and $\delta_v$ in \eqref{eq:deltav}. The first two counterterms, $\delta_z$ and $\delta_\alpha$, carry a scheme factor $c_z$ associated to the two-loop renormalization of the one-dimensional field $z$, as addressed in section \ref{app:Zz}. Thus, their contributions will depend on both $c_J$ and $c_z$. Once we include the counterterm factors, the remaining integral is exactly the one in \eqref{eq:oneloopintegral}. Therefore, we find
\begin{equation}
\begin{split}
     \vcenter{\hbox{\includegraphics[width=0.11\textwidth]{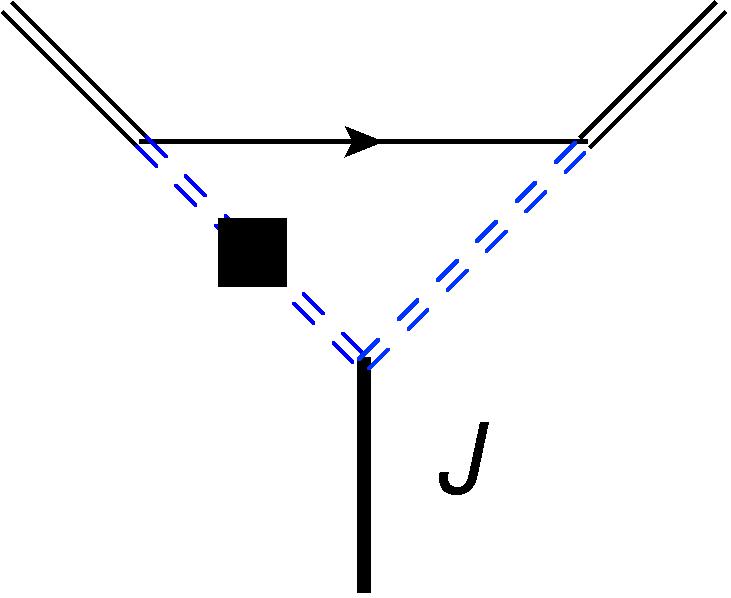}}} &=\vcenter{\hbox{\includegraphics[width=0.11\textwidth]{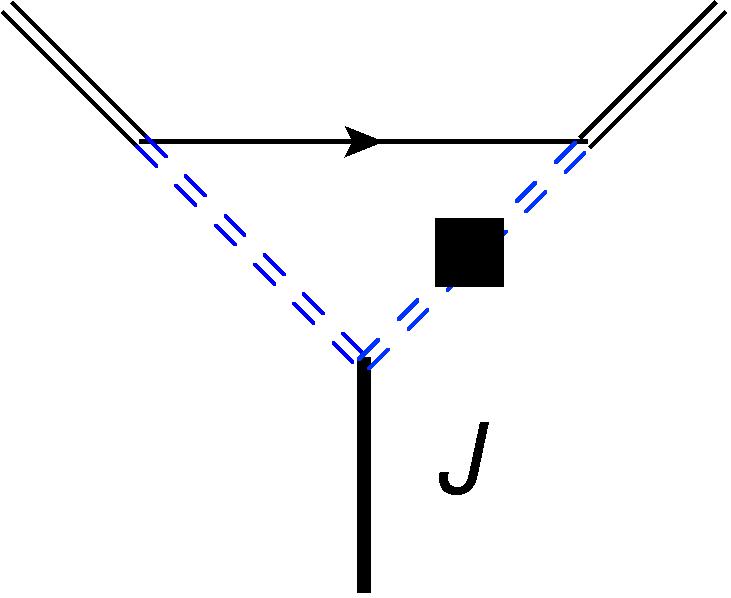}}}= -\bar\alpha^2\alpha^2 \frac{g^4N^2}{16\pi^2}\, \frac{1}{\epsilon} \, (c_z)^{\epsilon}(c_J)^{\epsilon}(\mu L)^{2\epsilon}\frac{\cos\frac{\theta}{2}}{\cos\frac{\varphi}{2}}\Bigg[ \frac{1}{\epsilon} + \log\left( 4\pi e^{\gamma_E} \right) \\ & \qquad\qquad\qquad\qquad\qquad\qquad\qquad\qquad - 2\log\left( 1+\sec\frac{\varphi}{2} \right) \Bigg] J\bar z z(0) \,,
     \end{split}
\end{equation}

\begin{equation}
    \begin{split}
     \vcenter{\hbox{\includegraphics[width=0.11\textwidth]{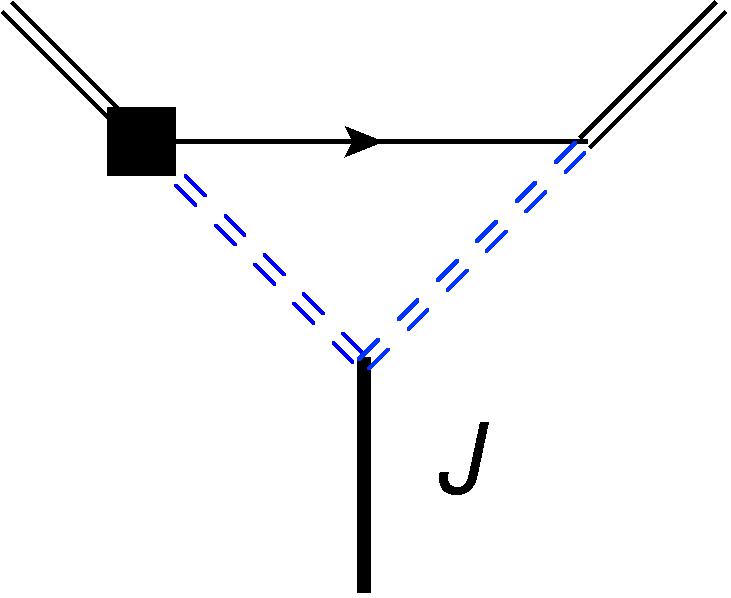}}} &=\vcenter{\hbox{\includegraphics[width=0.11\textwidth]{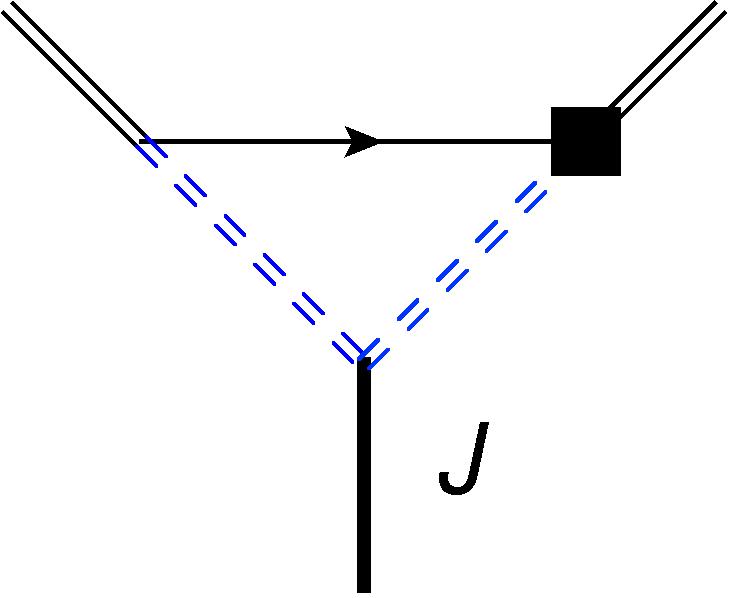}}}= \bar\alpha\alpha \frac{g^4N^2}{16\pi^2}\, \frac{1}{\epsilon} \, (c_z)^{\epsilon}(c_J)^{\epsilon}(\mu L)^{2\epsilon}\frac{\cos\frac{\theta}{2}}{\cos\frac{\varphi}{2}}\Bigg[ \frac{1}{\epsilon} + \log\left( 4\pi e^{\gamma_E} \right)\\ & \qquad\qquad\qquad\qquad\qquad\qquad\qquad\qquad - 2\log\left( 1+\sec\frac{\varphi}{2} \right) \Bigg] J\bar z z(0) \,,
    \end{split}
\end{equation}

\begin{equation} 
\begin{split}
     \vcenter{\hbox{\includegraphics[width=0.11\textwidth]{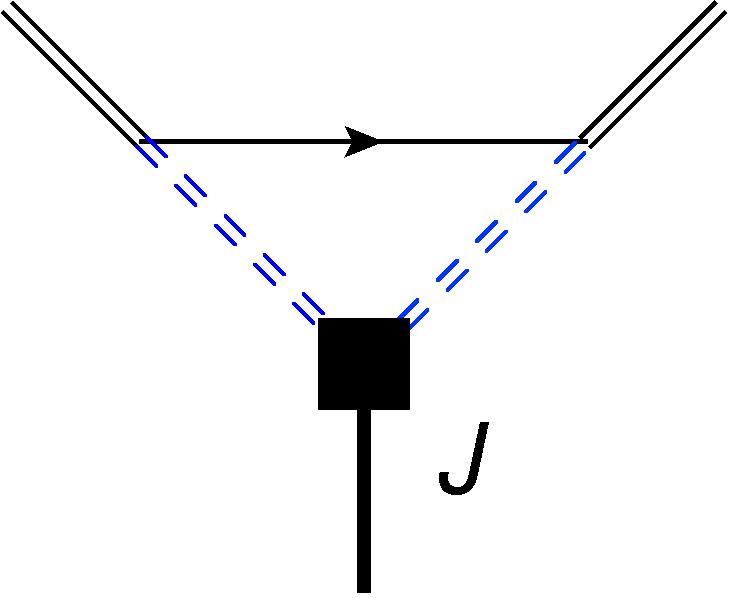}}} &= \bar\alpha^2\alpha^2 \frac{g^4N^2}{16\pi^2} \, \frac{1}{\epsilon} \, (c_J)^{2\epsilon}(\mu L)^{4\epsilon}  \left(\frac{\cos\frac{\theta}{2}}{\cos\frac{\varphi}{2}}\right)^2 \Bigg[ \frac{1}{\epsilon} +  \log\left(4\pi e^{\gamma_E} \right)\\ & \qquad\qquad\qquad\qquad\qquad\qquad\qquad\qquad -2\log\left(1+\sec\frac{\varphi}{2} \right)\Bigg] J\bar z z(0)\,.
\end{split}
\end{equation}

Collecting everything and adding a two-loop counterterm to remove these divergences, we are eventually led to the following renormalization function at two loops
\begin{equation}
\label{eq:Zv}
    \begin{split}
        Z_v =& 1- \bar\alpha\alpha\frac{g^2N}{4\pi}   \frac{\cos\frac{\theta}{2}}{\cos\frac{\varphi}{2}} \, \frac{1}{\epsilon} \, (c_J\mu^2L^2)^{\epsilon}  \\
        & + \bar\alpha\alpha\frac{g^4N^2}{32\pi^2\epsilon^2}(c_J \mu^2L^2)^{\epsilon}\frac{\cos\frac{\theta}{2}}{\cos\frac{\varphi}{2}}\left[ (c_J\mu^2 L^2)^{\epsilon}\left( \bar\alpha\alpha\frac{\cos\frac{\theta}{2}}{\cos\frac{\varphi}{2}} +2(\bar\alpha\alpha-1) \right) -4(c_z)^{\epsilon}(\bar\alpha\alpha-1 ) \right] \\ & - \frac{g^4N^2}{32\pi^2\epsilon} (c_J \mu^2L^2)^{\epsilon}\Bigg[ (c_z)^{\epsilon} 4\bar\alpha\alpha(\bar\alpha\alpha-1)\frac{\cos\frac{\theta}{2}}{\cos\frac{\varphi}{2}}\left[ \log\left( 4\pi e^{\gamma_E} \right) -2 \log\left( 1+ \sec\frac{\varphi}{2} \right) \right] \\ &\qquad+(c_J\mu^2 L^2)^{\epsilon}\Big[ 2(\bar\alpha^2\alpha^2-1)\varphi\cot\varphi - (\bar\alpha\alpha-1)\frac{\varphi}{\sin\varphi} \left( 1-\bar\alpha\alpha + (\bar\alpha\alpha+1)\cos\theta \right) \\ &\qquad\qquad - 4\bar\alpha\alpha\frac{\cos\frac{\theta}{2}}{\cos\frac{\varphi}{2}}\left[\bar\alpha\alpha\log\left( e^{\gamma_E}\pi \cos\frac{\varphi}{2}\sec^4\frac{\varphi}{4} \right) - \log\left( 4\pi e^{\gamma_E } \right) - 2\log\left(1+\sec\frac{\varphi}{2}\right)\right]\\ &\qquad\qquad-4\bar\alpha^2\alpha^2 \log\cos\frac{\varphi}{2}\Big] \Bigg]\,.
    \end{split}
\end{equation}

\subsection{Cusp anomalous dimension}
\label{sec:cuspanomalousdimension}

We now have all the ingredients to compute the cusp anomalous dimension and the corresponding Bremsstrahlung functions.

First of all,  exploiting the previous results we easily evaluate the cusp renormalization function $Z_{\rm cusp} = {Z_z}/{Z_v}$ at two loops. Before expanding \eqref{eq:Zz} and \eqref{eq:Zv} at small $\epsilon$, the full expression for $\log{Z_{\rm cusp}}$ reads 
\begin{equation}
    \label{eq:Zcusp}
\begin{split}
       \log & Z_{\text{cusp}}= \bar\alpha\alpha\frac{g^2N}{4\pi\epsilon} \left( \frac{\cos\frac{\theta}{2}}{\cos\frac{\varphi}{2}}(c_J\mu^2L^2)^\epsilon - (c_z)^\epsilon \right)  \\ 
       & \quad - \frac{g^4N^2}{16\pi^2\epsilon^2}\bar\alpha\alpha (\bar\alpha\alpha-1)\Bigg(\frac{\cos\frac{\theta}{2}}{\cos\frac{\varphi}{2}} \left((c_J\mu^2L^2)^{2\epsilon}-2(c_z c_J \nu^2L^2)^{\epsilon}\right)+ (c_z)^{2\epsilon}\Bigg) \, \\
       & \quad - \frac{g^4N^2}{32\pi^2\epsilon} \Bigg\{ 2 \bar\alpha\alpha(\bar\alpha\alpha -1) \left((c_z)^{2\epsilon} + 2 \frac{\cos\frac{\theta}{2}}{\cos\frac{\varphi}{2}} \log\left(4\pi e^{\gamma_E}\left(1+\sec\frac{\varphi}{2}\right)^{-2}\right) (c_z c_J\mu^2L^2)^{\epsilon}\right) \\
       & \quad +\Bigg[\frac{\bar\alpha\alpha-1}{\bar\alpha\alpha+1} \varphi \csc\varphi (\cos\theta -\bar\alpha^2\alpha^2+1) + 4\bar\alpha^2\alpha^2 \log\cos\frac{\theta}{2} - 2 \varphi \cot{\varphi}(\bar\alpha^2\alpha^2-1)\\
       & \quad+ 4\bar\alpha\alpha \frac{\cos\frac{\theta}{2}}{\cos\frac{\varphi}{2}} \left[\bar\alpha\alpha \log\left(\pi e^{-\gamma_E} \cos\frac{\varphi}{2}\sec^4\frac{\varphi}{4}\right)-\log\left(\frac{4\pi e^{\gamma_E}}{\left(1+\sec\frac{\varphi}{2}\right)^{2}} \right)\right]\Bigg](c_J\mu^2L^2)^{2\epsilon}\Bigg\} \,.
\end{split}
\end{equation}

This expression is manifestly scheme dependent, and  in a generic renormalization scheme it does not reproduce the expected result $Z_{\text{cusp}} = 1$ in the $\theta,\varphi \to 0$ limit (the trivial line). Rather, expanding in $\epsilon$ we are left with the following scheme dependent, finite expression
\begin{equation}\label{eq:limit1}
\begin{split}
    \lim_{\theta,\varphi\to 0} \log Z_{\text{cusp}} =\left[ \bar\alpha\alpha\frac{g^2N}{4\pi}-\bar\alpha\alpha(\bar\alpha\alpha-1) \frac{g^4N^2}{16\pi^2}\log\left(\frac{\pi^2e^{2\gamma_E-2}c_J\mu^2L^2}{c_z}\right)\right]\log\left( \frac{c_J \mu^2 L^2}{c_z} \right)  \,.
\end{split}
\end{equation}
Therefore, in order to restore $Z_{\text{cusp}}=1$ in the limit, we are forced to remove \eqref{eq:limit1} by a finite renormalization of $\log Z_{\text{cusp}}$\footnote{At the 1/6 BPS fixed point ($\bar\alpha\alpha=0$) scheme dependence disappears completely, whereas a scheme-dependent finite one-loop contribution survives at the 1/2 BPS fixed point ($\bar\alpha\alpha=1$). This does not contradict the results of \cite{Griguolo:2012iq, Bianchi:2017svd}, since in those papers it was implicitly assumed to work in a scheme where there were no residual finite contributions in the $\theta,\varphi \to 0$ limit.}.

At the $\bar\alpha\alpha = 1$ fixed point, expression \eqref{eq:Zcusp} reproduces $\log Z_{\rm cusp}$ for the cusped 1/2 BPS line \cite{Griguolo:2012iq}. In particular, the double pole appearing at two loops vanishes. This can be used as a perturbative argument in favor of the exponentiation of divergences as in \eqref{eq:exponentiation} (in dimensional regularization $\log{\mu/\Lambda}$ is replaced by $1/\epsilon$). In fact, we recall that while in four dimensions such an exponentialization is predicted by a solid theorem \cite{Dotsenko:1979wb,Gatheral:1983cz,Frenkel:1984pz}, in ABJM theory whether it occurs is still a challenging question. 
Our result \eqref{eq:Zcusp} shows that the double pole actually vanishes at both fixed points, $\bar\alpha\alpha = 1$ and $\bar\alpha=\alpha = 0$, supporting the conjecture that exponentiation should work not only for the cusped 1/2 BPS line, but also for 1/6 BPS bosonic one.

For generic $\bar\alpha\alpha$, instead, exponentiation does not work, as $\log Z_{\text{cusp}}$ exhibits a double pole which is not the square of the one-loop one. Since in general exponentation guarantees the finiteness of the cusp anomalous dimension, in this case a consistent result for $\Gamma_{\text{cusp}}$ is questionable. However, as we are now going to show, our $\Gamma_{\text{cusp}}$ is indeed two-loop finite.

The cusp anomalous dimension is defined following the non-standard prescription \eqref{eq:gammacuspL}. Taking into account that result \eqref{eq:Zcusp}, subtracted by \eqref{eq:limit1}, depends on the $\mu$ scale explicitly and through its dependence on $g$ and $\bar\alpha\alpha$, which are in turn functions of $\mu$ via their $\beta$-functions \eqref{eq:beta}, 
the cusp anomalous dimension evaluates to (we replace $g^2 \to \frac{2\pi}{k}$)
\label{eq:Gammacusp_res1}
\begin{align}   
\Gamma_{\text{cusp}}(\varphi, \theta, \bar\alpha, \alpha) =&\frac{N}{k}\bar\alpha\alpha \left( 1-\frac{\cos\frac{\theta}{2}}{\cos\frac{\varphi}{2}} \right) \\
&
+
    \frac{N^2}{k^2}\Bigg[ \left[(1-\bar\alpha^2\alpha^2)\left(2\cos\varphi - \cos\theta \right) -(\bar\alpha\alpha-1)^2 \right]\frac{\varphi}{2\sin\varphi} \nonumber \\ &\hskip 1.8cm +\bar\alpha\alpha(\bar\alpha\alpha-1)+2\bar\alpha^2\alpha^2\left(\frac{\cos\frac{\theta}{2}}{\cos\frac{\varphi}{2}} -1 \right)\log\sec\frac{\varphi}{2} \nonumber \\
    & \hskip 1.8cm 
+\bar\alpha\alpha(\bar\alpha\alpha-1)\left( \frac{\cos\frac{\theta}{2}}{\cos\frac{\varphi}{2}} - 1 \right)\log\left( \frac{c_J}{c_z} \mu^2L^2 \right)\Bigg]
    + \cO \left( \frac{N^3}{k^3} \right) \,. \nonumber 
\end{align}
This is a finite, well-defined cusp anomalous dimension for any defect theory along the RG flow, that is for any cusped $1/6$ BPS fermionic Wilson line. 

This quantity interpolates between the two cusp anomalous dimensions at the RG fixed points, the known $1/2$ BPS one for $\bar\alpha\alpha=1$ \cite{Griguolo:2012iq}, and the brand new result for the $1/6$ bosonic for $\bar\alpha\alpha=0$
\begin{equation}
    \Gamma_{\text{cusp}}^{\text{bos}} = \frac{N^2}{k^2}\frac{\varphi}{2\sin\varphi}\left( 2\cos\varphi - \cos\theta-1 \right) + \cO \left( \frac{N^3}{k^3} \right) \,.
\end{equation}

It is important to stress that the cusp anomalous dimension in \eqref{eq:Gammacusp_res1} is renormalization group invariant, \emph{i.e.} $\mu\frac{d}{d\mu}\Gamma_{\text{cusp}} = 0$. This is a strong consistency check of our definition \eqref{eq:gammacuspL} for the interpolating cusp anomalous dimension.

\section{The interpolating latitude}
\label{sec:latitudeWLs}

We now focus on the construction of parametric operators defined on a latitude circle, which should interpolate between the $1/6$ BPS fermionic latitude and the $1/12$ BPS bosonic one \cite{Bianchi:2014laa}. As already reviewed in the introduction, at the two fixed points latitude operators are known to be strictly related to cusped ones through the Bremsstrahlung function. We now want to investigate how this cusp/latitude correspondence gets modified in the presence of marginally relevant deformations. 

An operator interpolating between the $1/6$ BPS fermionic latitude and the $1/12$ BPS bosonic one has been previously introduced in \cite{Castiglioni:2022yes}. Here we resume its construction including a few more details. 

\begin{figure}[H]
    \centering
    \includegraphics[width=0.35\textwidth]{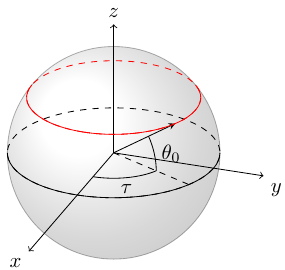}
\caption{In red, the latitude curve supporting the WL under investigation. The great circle corresponds to $\theta_0 = 0$.}
\label{fig:setup}
\end{figure}

We start by considering operators supported on the latitude circle 
\begin{equation}
\label{eq:latitudecurve}
    x^{\mu}(\theta_0,\tau) = (\sin\theta_0, \cos\theta_0\cos\tau,  \cos\theta_0\sin\tau)\,,
\end{equation}
where $\theta_0 \in [-\tfrac{\pi}{2}, \tfrac{\pi}{2}]$ is the (fixed) latitude angle, and $\tau$ parametrizes the contour.
The set-up is illustrated in figure \ref{fig:setup}. 

The latitude WL may carry also a dependence on an internal angle freely chosen in $[0, \tfrac{\pi}{2}]$, which rotates matter R-symmetry indices. In what follows we are not going to include it in the discussion.

As a starting point for the deformation we consider the bosonic $1/12$ BPS operator introduced in \cite{Bianchi:2014laa}, which is invariant under the action of the following two supercharges 
\begin{equation}
\label{eq:susycharges}
\begin{split}
{\cal Q}_1 = \sqrt{1+\nu}\left(\,Q_{12}^+\, - i e^{-i\theta_0} S_{12}^+\,\right) &- i\sqrt{1-\nu}\left(\,Q_{23}^-\,+ie^{-i\theta_0}S_{23}^-\,\right)\,,\\
{\cal Q}_2 = \sqrt{1+\nu}\left(Q_{34+} - i e^{i\theta_0} S_{34+}\right) &- i\sqrt{1-\nu}\left(Q_{14-}+ie^{i\theta_0}S_{14-}\right)\, ,
\end{split}
\end{equation}
where we have defined $\nu\equiv \cos\theta_0$.

There are in principle two operators preserving $\cQ_1$ and $\cQ_2$, which are separately charged under each node of the ABJ(M) quiver. We combine them in terms of a composite superconnection
\begin{equation}
\label{eq:bosonicL}
    \cL_0=\begin{pmatrix}
        \cA+\mathfrak{c} && 0 \\
        0 && \hat \cA 
    \end{pmatrix}\qquad \text{where}\qquad \begin{cases} \cA \equiv A_{\mu}\dot x^{\mu}-ig^2 |\dot x|M_J^{\ I}C_I \bar C^J\\
    \hat\cA \equiv A_{\mu}\dot x^{\mu}-ig^2 |\dot x|M_J^{\ I}\bar C^J C_I\\
\end{cases}\,,
\end{equation}
and
\begin{equation}
\label{eq:Mlatitudebos}
    M_J^{\ I} = \begin{pmatrix} -\nu && 0 && e^{-i\tau}\sqrt{1-\nu^2} && 0 \\
0 && -1 && 0 && 0 \\
e^{i\tau}\sqrt{1-\nu^2} && 0 && \nu && 0 \\
0 && 0 && 0 && 1
\end{pmatrix}\,.
\end{equation}

The factor $\mathfrak{c}$ in \eqref{eq:bosonicL} stands for a constant shift that is fixed by supersymmetry, as we will detail momentarily.

It is convenient to introduce the rotated scalar basis $(\tilde{C}_1,C_2,\tilde{C}_3,C_4)$,
\begin{equation}
    \begin{split}
        \tilde{C}_1 \equiv \sqrt{1+\nu}\, C_1 - \sqrt{1-\nu}\,e^{-i\tau} C_3\,, \qquad \tilde{C}_3 \equiv \sqrt{1+\nu}\, C_3 + \sqrt{1-\nu}\, e^{i\tau} C_1\,,
    \end{split}
\end{equation}
which diagonalizes the scalar coupling matrix \eqref{eq:Mlatitudebos} to $M=\diag(-1,-1,1,1)$. 

As done for the line, starting from the bosonic latitude we can construct a fermionic latitude preserving the same supercharges, by applying the ``hyperloop prescription'' \cite{Drukker:2019bev,Drukker:2020dvr}.
The fermionic superconnection is still defined as 
\begin{equation}\label{eq:superconnection2}
\cL = \cL_0 + i {\cal Q} G + G^2 \, ,     
\end{equation}
where now $\cQ$ is the sum of the two supercharges in \eqref{eq:susycharges} and $G$ is an off-diagonal matrix comprised of scalar fields. It is determined by the condition that under the action of $\cQ^2$ it transforms as a covariant derivative, where the covariant derivative includes the bosonic connections $\cA$ and $\hat\cA$ augmented by constant shifts (we take them to be always associated to 
the first node, as in \eqref{eq:bosonicL}). For details on the prescription we refer to \cite{Drukker:2020dvr}.

We find that the action of $\cQ^2$ on the scalar fields splits them into two families, according to the shift they are associated with. The relation between fields and shifts is presented in table \ref{table:fieldsandshifts}.
We see that while the non-rotated set (indices 2 and 4) is compatible with a $\nu$-dependent shift, the rotated set (indices 1 and 3) goes with a $\nu$-independent shift.

\begingroup

\setlength{\tabcolsep}{6pt} 
\renewcommand{\arraystretch}{1.3} 

\begin{table}[H]
\centering
\begin{tabular}{|c|c|}
\hline
\textbf{Fields in $G$}    & \textbf{Associated Shift} \\ \hline
$\quad \tilde{C}_1,\bar{\tilde{C}}^1\quad$ & $\quad +\frac{1}{2}\quad $  \\ \hline
$C_2,\bar{C}^2$ & $+\frac{\nu}{2}$  \\ \hline
$\tilde{C}_3,\bar{\tilde{C}}^3$ & $-\frac{1}{2}$  \\ \hline
$C_4,\bar{C}^4$ & $-\frac{\nu}{2}$  \\ \hline
\end{tabular}
\caption{Pairs of fields in $G$ and the corresponding shift $\mathfrak{c}$ to be added to the first node.}
\label{table:fieldsandshifts}
\end{table}
\endgroup

Although in principle the two elements of a given family (rotated or non-rotated) correspond to shifts with different signs, we can always perform a gauge transformation to bring them to be associated with the same shift. They can then be simultaneously included in $G$, however with a non trivial $\tau$ dependent relative phase.

For example, for the rotated pair this construction leads to the following form for $G$
\beq
\label{eq:Grot}
G=\sqrt{2i} g
\begin{pmatrix}
0 && \alpha_1 \bar{\tilde{C}}^1 + e^{i\tau} \bar\beta_3 \bar{\tilde{C}}^3 \\
\bar\alpha^1 \tilde{C}_1 + e^{-i\tau} \beta^3 \tilde{C}_3 && 0
\end{pmatrix}\,,
\eeq
where  $\bar\alpha^1,\alpha_1$ and $\beta^3,\bar\beta_3$ are complex (not complex conjugates) parameters. Therefore, they can be included simultaneously, thus parametrizing a single operator.

For the non-rotated pair, in principle $G$ shoud be
\beq
\label{eq:Gnonrot}
G=\sqrt{2i} g
\begin{pmatrix}
0 && \bar\alpha^2 C_2 + ie^{-i\nu \tau} \beta^4 C_4 \\
\alpha_2 \bar{C}^2 + ie^{i\nu \tau} \bar\beta_4 \bar{C}^4 && 0
\end{pmatrix}\,,
\eeq
where $\bar\alpha^2,\alpha_2$ and $\beta^4,\bar\beta_4$ are again complex (but not complex conjugates) parameters.
Nevertheless, an obstruction arises here, which prevents us from including both $C_2$ and $C_4$ in the same $G$. In fact, when $\cQ$ acts on the superconnection $\cL$ in \eqref{eq:superconnection2} it gives rise to the supercovariant derivative \cite{Drukker:2009hy,Lee:2010hk} of $G$,
\begin{equation}
    \cQ \cL = \mathfrak{D}_\tau G = \partial_\tau G + i \{\cL,G]\,. 
\end{equation}
This is nothing but a supergauge transformation that upon integration should leave the operator invariant. However, due to the $\nu$-dependent phases the supergauge transformation does not have definite boundary conditions. We are then forced to consider two separate branches of operators, one parameterized by $\bar\alpha^2,\alpha_2$ and one by $\beta^4,\bar\beta_4$.

In summary, for a $\nu$-independent shift $G$ we can include both $\bar\alpha^1,\alpha_1$ and $\beta^3, \bar\beta_3$ whereas for the $\nu$-dependent $G$ includes either $\bar\alpha^2,\alpha_2$ or $\beta^4,\bar\beta_4$. 

We could bypass such a subtlety by taking multiple copies of the nodes of the underlying theory and associating different shifts to different copies of the same node. This means taking a cover of the theory where some copies of the nodes will contain $\nu$-dependent shifts whereas others will contain $\nu$-independent ones.
We are not going to pursue this direction here. Rather,  we will consider the simplest setting where we do not take multiple copies and restrict to the study of 2-node operators. 

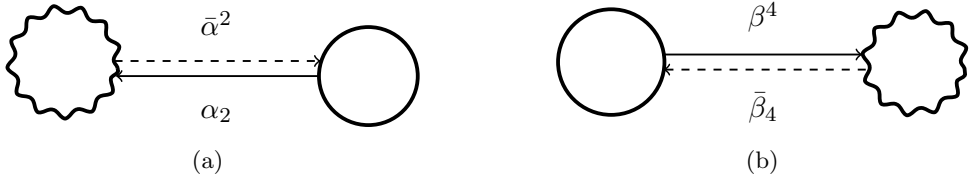
\begin{figure}[H]
    \centering
    \subfigure[]{
    \begin{tikzpicture}
\draw[decoration={snake,amplitude = .5mm,segment length=3.46mm},decorate, line width=.5mm] (6,2) circle (7mm);
\draw[line width=.5mm] (10,1.9) circle (6.5mm);

\draw[line width=.25mm,->,dashed] (6.66,2.1) to (9.37,2.1);

\draw[line width=.25mm,<-] (6.68,1.9) to (9.35,1.9);

\draw (6,2) node  []  {};
\draw (10,2) node  []  {};

\draw (8,2.6) node  []  {$\bar\alpha^2$};
\draw (8,1.4) node  []  {$\alpha_2$};

\end{tikzpicture}
\label{subfig:thetaneq0loops1}} \qquad\qquad
    \subfigure[]{\begin{tikzpicture}

\draw[line width=.5mm] (6,2) circle (7mm);
\draw[decoration={snake,amplitude = .5mm,segment length=3.46mm},decorate, line width=.5mm] (10,1.9) circle (6.5mm);

\draw[line width=.25mm, dashed,<-] (6.69,1.9) to (9.4,1.9);

\draw[line width=.25mm,->] (6.67,2.1) to (9.3,2.1);
\draw (6,2) node  []  {};
\draw (10,2) node  []  {};

\draw (8,1.4) node  []  {$\bar\beta_4$};
\draw (8,2.6) node  []  {$\beta^4$};
\end{tikzpicture}
\label{subfig:thetaneq0loops2}}
    \caption{Branches of $1/12$ BPS latitude loops. Points where supersymmetry is enhanced correspond to $\alpha_2\bar\alpha^2=-\bar\beta_4\beta^4=1$, where an $SU(2)$ subgroup of R-symmetry is restored and the operators become $1/6$ BPS.}
    \label{fig:thetaneq0loops}
\end{figure}

Precisely, we focus on the $\nu$-dependent case, which is the one distinguishing the latitude. We have two possible loops built out of
\begin{equation} 
\label{eq:Gmatrices}
G=\sqrt{2i} g 
\begin{pmatrix}
    0 && \bar\alpha^2 C_2 \\
    \alpha_2 \bar{C}^2 && 0
\end{pmatrix}\qquad\text{or}\qquad G 
=\sqrt{2i} g
\begin{pmatrix}
    0 && \beta^4 C_4 \\
    \bar\beta_4 \bar{C}^4 && 0
\end{pmatrix}\,.    
\end{equation}
If we want to avoid the appearance of inconvenient phases in $G$,  for each option there is a precise choice of the node that should accomodate the constant shift. These are represented in figure \ref{fig:thetaneq0loops}, where the squigglyness indicates the node with constant shift $\frac{\nu}{2}$. Inverting the position of the squigglyness in each figure would correspond to adding phases to $\alpha^2,\bar\alpha_2$ or $\beta^4,\bar\beta_4$ (see for instance eq. \eqref{eq:Gnonrot}).

For concreteness, we will focus on the operator in figure \ref{subfig:thetaneq0loops1}, that is the supertraced holonomy of a superconnection that is supplemented by a constant shift placed in the first node and has periodic off-diagonal elements. 

It is always possible to apply a gauge transformation to remove the constant shift in the superconnection at the price of adding a twist matrix $\cT$ and changing the periodicity of the off-diagonal elements. Details can be found in \cite{Castiglioni:2022yes}. 
Since it turns out that the use of shifted connections is not very convenient for performing perturbative analysis, here we will use the formulation with the $\cT$ matrix, which in this case reads
\begin{equation}
    \cT \equiv \begin{pmatrix}
    \mathds{1}_N e^{-\frac{i\pi\nu}{2}} && 0 \\
    0 && \mathds{1}_N e^{\frac{i\pi\nu}{2}}
\end{pmatrix}\,.
\end{equation}
In this set-up the operator can be explicitely written as
\begin{equation}
\label{eq:latitudeWL}
    W_\nu=\frac{1}{\left(\sTr \cT\right)} \, \sTr\cP \bigg(e^{-i\oint \cL \,d\tau}\cT\bigg)\,,\qquad \cL = \begin{pmatrix}
        \cA && \bar{f}\\
        f && \hat{\cA}
    \end{pmatrix}
\end{equation}
with $\cA$ and $\hat\cA$ defined as in \eqref{eq:bosonicL}, with scalar coupling matrix in the $(C_1,C_2,C_3,C_4)$ basis given by (from now on we rename
$\alpha_2 \equiv \alpha$, $\bar\alpha^2 \equiv \bar\alpha$)
\begin{equation}
\label{eq:Mlatitude}
    M_J^{\ I} = \begin{pmatrix} -\nu && 0 && e^{-i\tau}\sqrt{1-\nu^2} && 0 \\
0 && -1 + 2 \bar\alpha\alpha && 0 && 0 \\
e^{i\tau}\sqrt{1-\nu^2} && 0 && \nu && 0 \\
0 && 0 && 0 && 1
\end{pmatrix}\,,
\end{equation}
and off-diagonal elements 
\begin{equation}
\label{eq:ffbar2node}
\begin{split}
 \bar{f} &= -\bar\alpha \, e^{\frac{i\nu\tau}{2}}\eta \, \bigg(\sqrt{\tfrac{1+\nu}{2}}\,\bar\psi^1-\sqrt{\tfrac{1-\nu}{2}}\,e^{i\tau}\bar\psi^3\bigg)\,, \\
 f &= -\alpha \, e^{-\frac{i\nu\tau}{2}} \xi \,  \bigg(  \sqrt{\tfrac{1+\nu}{2}}\,\psi_1 - \sqrt{\tfrac{1-\nu}{2}}\, e^{-i\tau}\psi_3\bigg)\,,
\end{split}
\end{equation}
where the commuting spinors $\eta$ and $\xi$ are 
\begin{equation}
\label{eq:spinorialcouplings}
\begin{split}
    \eta^\delta = g\sqrt{i}(1,-ie^{-i\tau})^\delta\,,\qquad\qquad\qquad \eta_\delta = g\sqrt{i}\begin{pmatrix} ie^{-i\tau} \\ 1 \end{pmatrix}_\delta\,, \\ \xi^\delta = g\sqrt{i}(-ie^{i\tau},1)^\delta\,,\qquad\qquad\qquad \xi_\delta = g\sqrt{i}\begin{pmatrix} -1 \\ -ie^{i\tau} \end{pmatrix}_\delta\,.
\end{split}
\end{equation}

This is a parametric latitude 
describing a 1/12 BPS fermionic Wilson loop. Upon varying $\bar\alpha \alpha$, it  interpolates between the 1/12 bosonic latitude in (\ref{eq:bosonicL}, \ref{eq:Mlatitudebos}) corresponding to $\bar\alpha \alpha = 0$, and the 1/6 BPS fermionic latitude \cite{Bianchi:2014laa} corresponding to $\bar\alpha \alpha = 1$.
Generalizing to the $\nu \neq 1$ case the analysis done in \cite{Castiglioni:2022yes} for parametric WLs defined on the maximal circle, the RG flows connecting the two fixed points are {\em enriched flows} made by a sequence of BPS but non-conformal points. The $\bar\alpha \alpha$ terms in (\ref{eq:Mlatitude}, \ref{eq:ffbar2node}) describe a marginally relevant deformation of the 1/12 BPS latitude WL.

\subsection{Perturbative renormalization}

Following \cite{Castiglioni:2022yes,Castiglioni:2023uus}, we study quantum properties of the parametric latitude  defined above, using the one-dimensional auxiliary theory approach reviewed in section \ref{app:1dmethod}. 

The perturbative evaluation of the latitude VEV mostly follows the one for the $1/6$ BPS fermionic latitude done in \cite{Bianchi:2014laa}. The only modifications are the appearence of $\bar\alpha\alpha$  prefactors in front of some of the integrals, and the need to keep track of the $O(\epsilon)$ term at one loop, as we are going to explain. Thus we rely on previous calculations and some refinements addressed in the following.

\begin{wrapfigure}{r}{0.3\textwidth}
    \centering
    \subfigure[]{\includegraphics[width=0.07\textwidth]{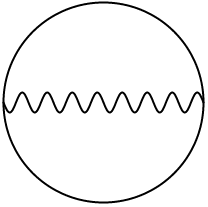}\label{subfig:oneloopdiagrama}} \hspace{8mm}
    \subfigure[]{\includegraphics[width=0.07\textwidth]{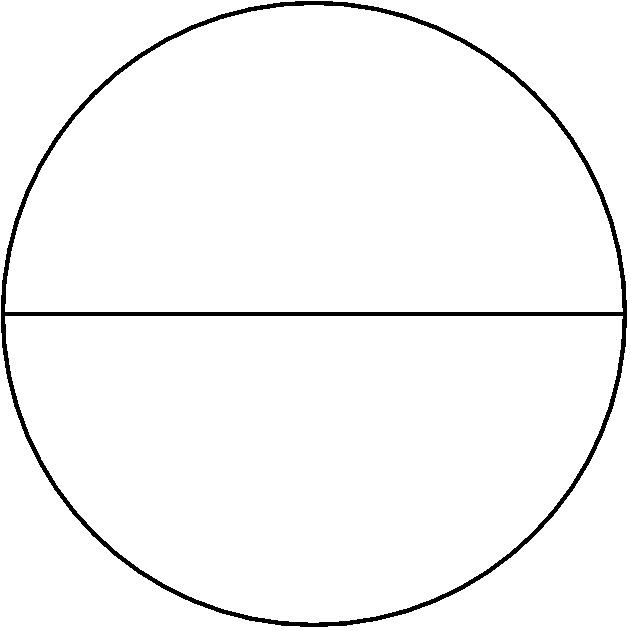}\label{subfig:oneloopdiagramb}}
    \caption{One-loop diagrams: single \subref{subfig:oneloopdiagrama} gauge and \subref{subfig:oneloopdiagramb} fermion exchanges.}
    \label{fig:oneloopdiagrams}
\end{wrapfigure}

At one loop there are two contributions coming from the exchange of a gauge field and a fermion field, see figure \ref{fig:oneloopdiagrams}. As is well known \cite{Bianchi:2013zda}, the one in figure \ref{subfig:oneloopdiagrama} is framing dependent and at framing zero it vanishes. On the other hand, the second one is always non-zero. Its finite part was computed in \cite{Bianchi:2014laa}
and reads 
\begin{equation}
\label{eq:finite}
\langle W_{\nu}\rangle_0^{(1)} =  \bar\alpha_0 \alpha_0 \, \frac{g^2N}{2} \nu \cot{\frac{\pi\nu}{2}} + {\cal O}(\epsilon).
\end{equation}

However, for the scope of the present calculation we need to keep track also of the ${\cal O}(\epsilon)$ term. in fact, this may give finite contributions at two loops once the parameters are renormalized. Since the evaluation of the ${\cal O}(\epsilon)$ term has not been done before, here we provide a few details of the calculation. 

First, starting from the integral corresponding to diagram \ref{subfig:oneloopdiagramb} 
\begin{equation}
-\int_0^{2\pi} d\tau_1 \int_0^{\tau_1} d\tau_2 \; \left( e^{-\frac{i\pi\nu}{2}}\langle \bar f(\tau_1)f(\tau_2) \rangle
- e^{\frac{i\pi\nu}{2}}\langle  f(\tau_1) \bar f(\tau_2) \rangle \right)\,,
\end{equation}
we  write 
\begin{equation}\label{eqn:lat1}
    \langle \bar f(\tau_1)f(\tau_2) \rangle =  g^2 (\mu R \cos\theta_0 )^{2\epsilon} N^2  \, \bar\alpha_0\alpha_0 \,  \frac{\Gamma(\frac{1}{2}-\epsilon)}{4^{1-\epsilon}\pi^{\frac{3}{2}-\epsilon}}\left(\frac{d}{d\tau_1} {\rm g}_{\epsilon}(\tau_{12})-i\epsilon 
    \nu\, {\rm g}_{\epsilon}(\tau_{12})\right)\,,
\end{equation}
where we have defined
\begin{equation}
\label{eq:gfunction}
{\rm g}_\epsilon(\tau) = \frac{e^{\frac{i\nu\tau}{2}}}{\left( \sin^2\frac{\tau}{2} \right)^{\frac{1}{2}-\epsilon}}\,.
\end{equation}
It follows that the ${\cal O}(\epsilon)$ contributions can arise from the evaluation of the following two integrals
\begin{equation}
    \cI_1 \equiv \int_0^{2\pi}d\tau_1 \int_0^{\tau_1}d\tau_2 \frac{d}{d\tau_1}{\rm g}_{\epsilon}(\tau_{12})\,, \quad \cI_2\equiv\int_0^{2\pi}d\tau_1 \int_0^{\tau_1}d\tau_2\, {\rm g}_{\epsilon}(\tau_{12})\,.
\end{equation}
We can rewrite the first one as
\begin{equation}
    \cI_1 = -\int_0^{2\pi} d\tau_1 \int_{0}^{\tau_1} d\tau_2 \frac{d}{d\tau_2} {\rm g}_{\epsilon}(\tau_{12}) = \int_0^{2\pi} d\tau_1 \big( {\rm g}_{\epsilon}(\tau_1) -{\rm g}_{\epsilon}(0) \big)\,,
\end{equation}
and follow the standard way of regularizing the result by discarding the ${\rm g}_\epsilon(0)$ term. Now, the two integrals can be easily computed by writing the denominator of ${\rm g}_\epsilon$ in \eqref{eq:gfunction} in terms of exponential functions (see \cite{Bianchi:2014laa} for details).

Since $\langle f(\tau_1)\bar f(\tau_2) \rangle$ is simply obtained from the previous result with the substitution $\nu \to-\nu$, summing up the two contributions, inserting the normalization factor $\cR$ and combining with the finite part \eqref{eq:finite} we obtain\footnote{Here the superscript stands for the loop order, whereas the subscript indicates that the result is expressed in terms of the bare parameters. 
}
\begin{equation}
\label{eqn:1loopbarelatitude}
    \langle W_\nu \rangle_0^{(1)}= \bar\alpha_0\alpha_0 \,  \, \frac{g^2N}{2}(R\nu  )^{2\epsilon} \nu \cot\tfrac{\pi\nu}{2} \left[ 1 + \epsilon\left(  \log (4\pi e^{\gamma_E})-H_{-\frac{1-\nu}{2}} - H_{-\frac{1+\nu}{2}}  \right) \right]\,,
\end{equation}
where $H_x$ are the harmonic numbers and we have restored the sphere radius $R$ for dimensional reasons. 

Differently from the case of the maximal circle \cite{Castiglioni:2022yes,Castiglioni:2023uus}, now the one-loop VEV is finite and no longer vanishing in the $\epsilon\to 0$ limit. This implies that, once we replace the bare parameters with the renormalized ones, it acquires divergent contributions, as well as finite (scheme dependent) terms present in the renormalization functions. Precisely,
expression \eqref{eqn:1loopbarelatitude} given in terms of the renormalized $\bar\alpha \alpha$ reads 
\begin{equation}
\label{eqn:1loop_ren}
\begin{split}
     \langle W_\nu \rangle^{(1)}  =&  \bar\alpha\alpha \frac{g^2N}{2} \nu \cot\tfrac{\pi\nu}{2} \\
    & +  \frac{\bar\alpha\alpha
    (\bar\alpha\alpha-1)}{4\pi} g^4N^2 (\mu R \nu  )^{2\epsilon} \nu \cot\tfrac{\pi\nu}{2} 
    \left( \frac{1}{\epsilon} +\log (4\pi e^{\gamma_E} c_z) - H_{-\frac{1-\nu}{2}} - H_{-\frac{1+\nu}{2}}\right).
\end{split} 
\end{equation}

At two loops it is convenient to distinguish between purely bosonic and purely fermionic diagrams, as the $\alpha, \bar\alpha$ parameters end up contributing only to the fermionic ones. In any case, one can easily check that the presence of the parameters does not affect the struture of the Feynman integrals, therefore the results found in \cite{Bianchi:2014laa} still hold and we simply need to adapt the coefficients accordingly.

In particular, in the planar limit the bosonic diagrams evaluate to
\begin{align}
    &\vcenter{\hbox{\includegraphics[width=0.07\textwidth]{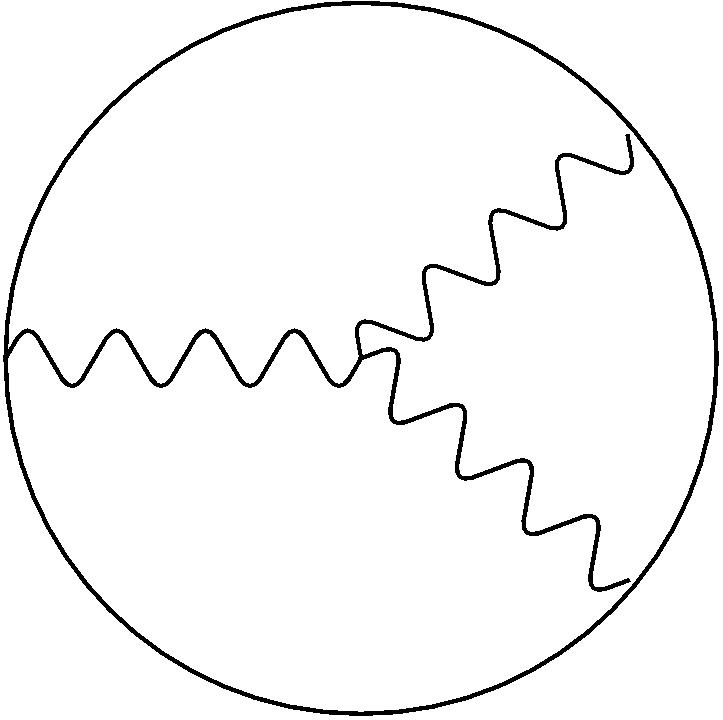}}} = 
    \frac{i g^4(\mu R\cos\theta_0)^{4\epsilon}}{12} N^3 \sin\frac{\pi\nu}{2}  \,, \\
    &\vcenter{\hbox{\includegraphics[width=0.07\textwidth]{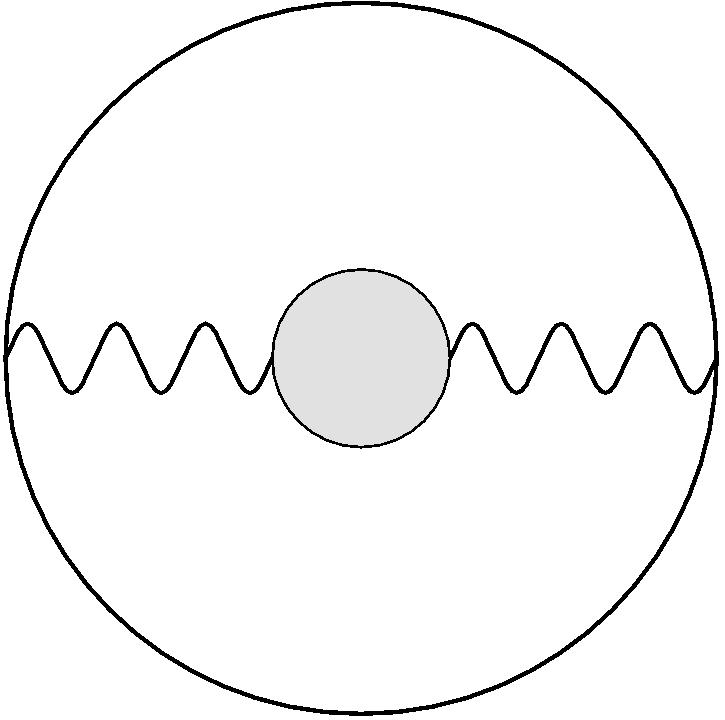}}}\,\, +\vcenter{\hbox{ \includegraphics[width=0.07\textwidth]{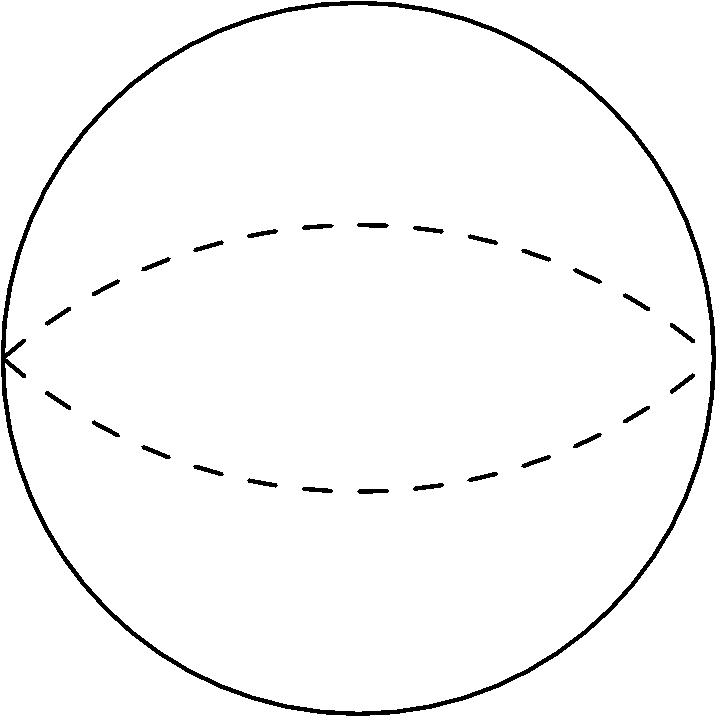}}} = -i\frac{g^4 (\mu R\cos\theta_0)^{4\epsilon}}{4} N^3(1+\nu^2)\sin\frac{\pi\nu}{2}   \,,
\end{align}
while the fermionic ones give 
\begin{equation}
\begin{split}
\vcenter{\hbox{\includegraphics[width=0.07\textwidth]{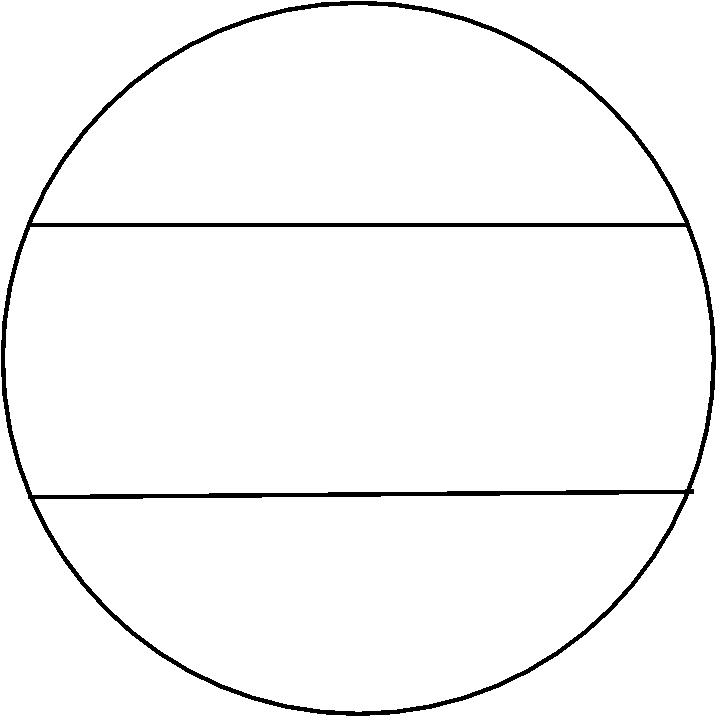}}} 
&= g^4 (\mu R \cos\theta_0)^{4\epsilon}(\bar\alpha_0\alpha_0)^2 \frac{N^3}{4}\nu\bigg\{i(\nu-4)\sin\tfrac{\pi\nu}{2} \\ &\hspace{40mm}+ \frac{2}{\pi}  \bigg(\frac{1}{\epsilon} -4H_{\frac{\nu-1}{2}} + 2\log(4\pi e^{\gamma_E} ) \bigg) \cos\tfrac{\pi\nu}{2}  \bigg\} \,,\\
\vcenter{\hbox{\includegraphics[width=0.07\textwidth]{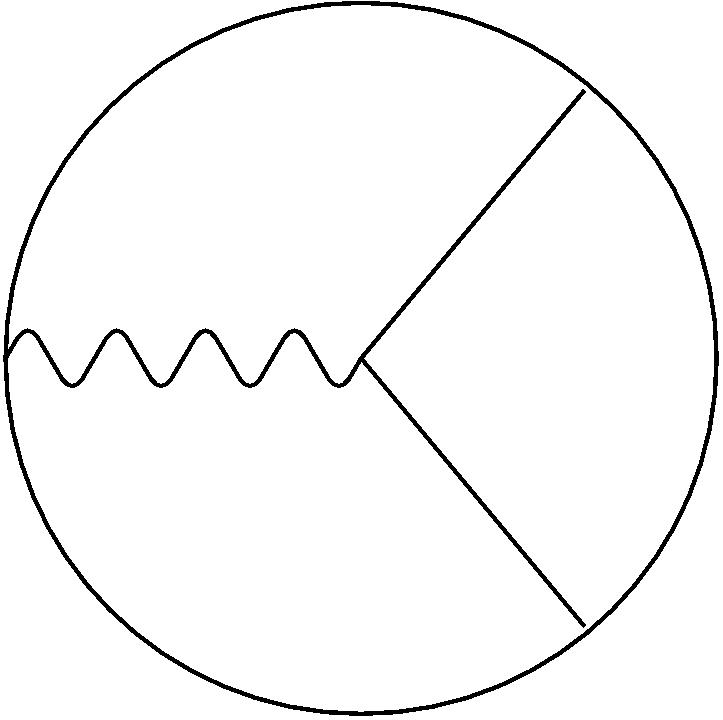}}} 
&=  g^4 (\mu R\cos\theta_0)^{4\epsilon}(\bar\alpha_0\alpha_0) \frac{N^3}{8}\nu \bigg\{\frac{4i}{\pi}  \bigg(\frac{1}{\epsilon} -4H_{\frac{\nu-1}{2}} + 2\log(4\pi e^{\gamma_E} ) \bigg)\cos\tfrac{\pi\nu}{2} -8  \sin\tfrac{\pi\nu}{2}\bigg\}.
\end{split}
\end{equation}
Including also the normalization factor $1/\sTr(\cT)$, the complete two-loop contribution sums up to 
\begin{equation}\label{eqn:2loop_res}
\begin{split}
    \langle W_\nu \rangle^{(2)} = -\frac{g^4N^2}{2}(\mu R \nu  )^{4\epsilon} \,\bigg\{&\frac{\bar\alpha\alpha(\bar\alpha\alpha-1)}{2\pi} \bigg(\frac{1}{\epsilon}+ 2\log(4\pi e^{\gamma_E})-4H_{\frac{\nu-1}{2}} \bigg) \, \nu\cot\tfrac{\pi\nu}{2}\\ &+ \left(
        \frac{1}{6} + \bar\alpha\alpha(\bar\alpha\alpha-1) \nu -\frac{(\bar\alpha^2\alpha^2-1)}{4} \nu^2 \right)\bigg\}\,.
\end{split}
\end{equation}
In principle, this expression is divergent in the $\epsilon \to 0$ limit. 
However, the renormalization at one loop has produced a divergent term in \eqref{eqn:1loop_ren} which removes exactly this divergence. 
It follows that  the renormalized VEV for the $1/12$ BPS parametric latitude Wilson loop, up to two loops, is finite and given by (again, $g^2 \to \frac{2\pi}{k}$) 
\begin{align}
\label{eq:finalWlatitude}
    \langle W_\nu \rangle= &1 -\frac{N}{k}\pi\bar\alpha\alpha \, \nu \cot\frac{\pi\nu}{2}  + \frac{N^2}{k^2}\frac{\pi}{6} \bigg[ -\pi \Big( 
       3\nu^2(\bar\alpha^2\alpha^2-1) -12\nu(\bar\alpha\alpha-1)\bar\alpha\alpha -2\Big)\\&+6\bar\alpha\alpha(\bar\alpha\alpha-1) \left( \log\left( \frac{4\pi e^{\gamma_E} \mu^2R^2\nu^2}{c_z} \right)+H_{-\frac{1+\nu}{2}}-3H_{\frac{\nu-1}{2}}\right) \nu \cot\frac{\pi\nu}{2} \bigg] +  \cO \left( \frac{N^3}{k^3} \right) \,. \nonumber
\end{align}
Using the one-loop $\beta$-functions in \eqref{eq:beta} 
it can be rewritten as 
\begin{align}
    \langle W_\nu \rangle= &1 -\frac{N}{k}\pi\bar\alpha\alpha \, \nu \cot\frac{\pi\nu}{2}  - \frac{N^2}{k^2}\frac{\pi^2}{6} \left( 
3\nu^2(\bar\alpha^2\alpha^2-1) -2\right)\\&+  \beta_{\bar\alpha \alpha} \, \frac{N}{k}\frac{\pi}{2} \left[ \left(\log\left( \frac{4\pi e^{\gamma_E}\mu^2R^2\nu^2}{c_z} \right)+H_{-\frac{1+\nu}{2}}-3H_{\frac{\nu-1}{2}}\right) \nu \cot\frac{\pi\nu}{2}  - 2 \nu \right] +  \cO \left( \frac{N^3}{k^3} \right) \,. \nonumber
\end{align}

As a consistency check it is easy to see that this expression satisfies the Callan-Symanzik equation
\begin{equation}
    \left( \mu \frac{\partial}{\partial \mu} +  \beta_{\bar\alpha\alpha} \frac{\partial}{\partial \bar\alpha\alpha} \right) \langle W_\nu \rangle = 0\,.
\end{equation}
Moreover, at the $\bar\alpha\alpha = 0,1$ fixed points it correctly reproduces the large $N$ limit of the 1/12 BPS bosonic and 1/6 BPS fermionic latitudes \cite{Bianchi:2014laa}.

Going back to \eqref{eqn:2loop_res}, we note that $1/\epsilon$ pole cancels in the $\nu\to 1$ limit, in agreement with the finiteness of the two-loop result for the maximal circle, previously found in \cite{Castiglioni:2022yes}. However, also in the $\nu= 1$ case there is no reason to expect finite loop contributions at any order, as long as the VEV is expressed in terms of bare parameters. Simply, the pattern we are obtaining at two loops for the latitude will appear at higher orders for the maximal circle. Of course, the same reasoning applies to the scheme dependent terms. This pattern is similar to what occurs for the 
interpolating latitude WL in ${\cal N}=4$ SYM \cite{Beccaria:2018ocq} and for WLs in non-conformal ${\cal N}=2$ SYM in four dimensions \cite{Billo:2023igr}.  

\section{The interpolating Bremsstrahlung functions}
\label{sec:interpolatingBfunction}

The coefficients of the cusp anomalous dimension in the small angles expansion are known as \emph{Bremsstrahlung functions} $B$. Precisely, they are defined as 
\begin{equation}\label{eq:ThetaPhi}
    \Gamma_{\text{cusp}} \sim \theta^2 B^{\theta} - \varphi^2 B^{\varphi}\,.
\end{equation}
In the introduction we have recalled their physical meaning, and  their relationship in the case of 1/2 BPS fermionic and 1/6 BPS bosonic operators in ABJM theory. 

An additional important property of these fuctions is that at the $1/2$ and $1/6$ BPS fixed points exact identities hold, which allow to express them as derivatives of circular Wilson loops. Originally introduced in \cite{Lewkowycz_2014} to express $B^{\varphi}_{1/6}$ as the derivative of a multi-winding circular Wilson loop with respect to the winding number $m$, 
\begin{equation}
    B^{\varphi}_{1/6} = \frac{1}{4\pi^2} \frac{\partial}{\partial m} \log |\langle W_m \rangle| \bigg|_{m=1}\,,
\end{equation}
this identity was later generalized to obtain $B_{1/2}$, $B^{\theta}_{1/6}$ and $B^{\varphi}_{1/6}$ as derivatives of {\em latitude} Wilson loops with respect to the latitude parameter $\nu$ \cite{Bianchi:2014laa, Bianchi:2017ozk, Aguilera_Damia_2014}
\begin{equation}\label{eq:generalidentity}
    B_{1/2} = \frac{1}{4\pi^2} \frac{\partial}{\partial \nu}\log |\langle 
    W_{1/6} \rangle| \bigg |_{\nu=1}\,, \qquad B_{1/6}^{\theta} = \frac12 B^{\varphi}_{1/6}= \frac{1}{4\pi^2} \frac{\partial}{\partial \nu}\log |\langle 
    W_{1/12} \rangle |\bigg |_{\nu=1}\,.
\end{equation}
The proof of these identities strongly relies on the (super)conformal invariance of the defect theory living on the Wilson loop. Therefore, outside the fixed points we do not expect them to be valid. Nevertheless, 
we want to study how they get modified when the superconformal defect at the fixed point is perturbed by a marginally relevant operator. 
Having computed $\Gamma_{\rm cusp}$ and the latitude VEV in the presence of parametric deformations, we have all the ingredients 
to address this question.

Expanding the two-loop result \eqref{eq:Gammacusp_res1} at small $\theta , \varphi$ angles we can easily extract the Bremsstrahlung functions at this order. For a generic interpolating operator we obtain two different functions
(restoring $g^2 \rightarrow 2\pi/k$)
\begin{equation}
\begin{split}\label{eq:Bfinal}
    B^{\theta} (\bar\alpha\alpha)&=  \frac18 \frac{N}{k} \bar\alpha\alpha - \frac14\frac{N^2}{k^2} (\bar\alpha^2\alpha^2-1) - \frac{1}{16} \frac{N}{k}\beta_{\bar\alpha\alpha}\log\left(\frac{c_J}{c_z} \mu^2 L^2 \right) \,,\\
    B^{\varphi} (\bar\alpha\alpha) &= \frac18 \frac{N}{k} \bar\alpha\alpha - \frac12 \frac{N^2}{k^2} (\bar\alpha^2\alpha^2-1) - \frac{1}{16} \frac{N}{k}\beta_{\bar\alpha\alpha}\log\left( 
    \frac{c_J}{c_z}  \mu^2 L^2 e^{-\frac{4}{3}} \right) \, ,
\end{split}
\end{equation}
where we have used the $\beta$-functions in \eqref{eq:beta} at $\epsilon =0$.

These results reproduce correctly the relations holding at the fixed points. In fact, setting $\bar\alpha \alpha = 0$ we obtain $B^{\theta} (0) = \tfrac12 B^{\varphi} (0)$, in agreement with the general identity \cite{Bianchi:2018scb}. Setting instead $\bar\alpha \alpha = 1$ we recover the well known result for the $1/2$ BPS case where the two functions coincide \cite{Griguolo:2012iq}.  

\begin{figure}
    \centering
    \includegraphics[width=0.8\textwidth]{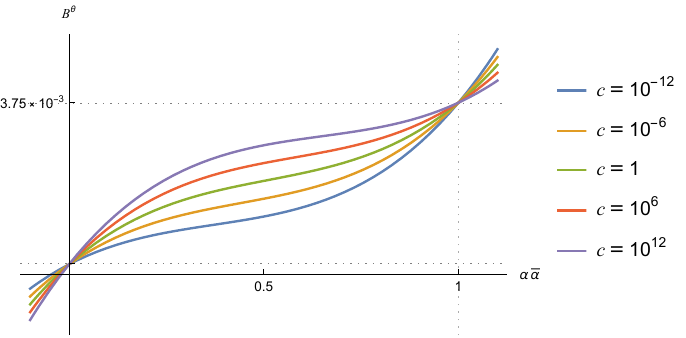}
    \caption{The interpolating Bremsstrahlung function $B^{\theta}$ for $\bar\alpha\alpha \in [0,1]$, $N/k=0.03$ and different values of $c = (\Lambda L)^2c_J/c_z$.}
\label{fig:Bremsstrahlung}
\end{figure}

Results \eqref{eq:Bfinal} clarify the structure of the RG flows for the Bremsstrahlung functions. The interpolating $B$'s have corrections at both even and odd orders in $N/k$ with a precise $\bar\alpha \alpha$ dependence: Odd orders are proportional to $\bar\alpha \alpha$, whereas  even orders come with a $(\bar\alpha \alpha -1)$ factor. Therefore they correctly interpolate between an odd function of $N/k$ at the 1/2 BPS fixed point, and an even function at the 1/6 BPS fixed point.

Inverting \eqref{eq:beta_solution}, we can trade $\mu$ in \eqref{eq:Bfinal} with its expression in terms of $\bar\alpha\alpha$. This imports the $\Lambda$ parameter in $B^{\theta}$, $B^{\varphi}$ such that they become eventually functions of the $\Lambda L$ combination. In figure \ref{fig:Bremsstrahlung} we plot $B^{\theta}$ for different values of the scheme parameter $c \equiv (\Lambda L)^2 c_J/c_z$.

\vskip 10pt

We now move to study how identities \eqref{eq:generalidentity} generalize away from the fixed points. In section \ref{sec:latitudeWLs} we computed the expectation value of the $1/12$ BPS latitude Wilson loop, which in the $\nu\to 1$ limit corresponds to the interpolating $1/6$ BPS fermionic operator.
When we take the derivative of the logarithm of $\langle W \rangle$ in \eqref{eq:finalWlatitude} with respect to $\nu$ we obtain (restoring $g^2=2\pi/k$)
\begin{equation}\label{eq:bremsstrahlung_latitude1}
   \frac{1}{4\pi^2} \frac{\partial \log \langle 
    W_\nu \rangle}{\partial \nu}\bigg|_{\nu=1} = \frac{1}{8}\frac{N}{k} \bar\alpha\alpha - \frac{1}{4}\frac{N^2}{k^2} (\bar\alpha^2\alpha^2-1) - \frac{1}{16}\frac{N}{k} \beta_{\bar\alpha\alpha}\log\left(\frac{4\pi }{c_z} e^{\gamma_E-2} \mu^2 R^2\right)\,.
\end{equation}
This expression reproduces correctly the values of $(B^{\theta}_{1/6}, B^{\varphi}_{1/6})$ and $B_{1/2}$, at $\bar\alpha \alpha = 0$ and $\bar\alpha \alpha 
 = 1$ fixed points respectively, in agreement with \eqref{eq:generalidentity}. 
 
For generic $\bar\alpha\alpha$, instead, comparing this result with $ B^{\theta}(\bar\alpha\alpha) $  and $ B^{\varphi}(\bar\alpha\alpha) $ in \eqref{eq:Bfinal} and neglecting higher order terms we are led to
\begin{align} \label{eq:Blatitude}
 B^{\theta}(\bar\alpha\alpha) &=  \frac{1}{4\pi^2} \frac{\partial \log \langle 
    W_\nu \rangle}{\partial \nu}\bigg|_{\nu=1} +   
    \beta_{\bar\alpha\alpha} \, \frac{1}{16}\frac{N}{k} \, \log\left(\frac{4\pi }{c_J}e^{\gamma_E-2}\frac{R^2}{L^2}\right)  \, ,  \\
    B^{\varphi} (\bar\alpha \alpha) &= \frac{1}{4\pi^2} \frac{\partial\log\langle W_\nu \rangle}{\partial \nu} \bigg|_{\nu=1} \!- \! \frac{1}{4}\frac{N^2}{k^2} (\bar\alpha^2\alpha^2-1) + \beta_{\bar\alpha\alpha}\frac{1}{16}\frac{N}{k}\log\left( \frac{4\pi }{c_J} e^{\gamma_E -\frac{2}{3}}\frac{R^2}{L^2} \right)  \nonumber \, .
\end{align}
The second term in the r.h.s of $ B^{\varphi}(\bar\alpha\alpha) $ accounts for the deviation of $B^{\varphi}$ from $B^\theta$ when moving from the 1/2 BPS to the 1/6 BPS fixed point. 
As expected, the general identities in \eqref{eq:generalidentity} are broken by contributions proportional to the conformal anomaly $\beta_{\bar\alpha\alpha}$, which introduces a (scheme-dependent) deviation starting at two loops. 

\section{Defect correlation functions}
\label{sec:defectcorrelationfunction}

(Super)conformal Wilson loops describe one-dimensional defects and provide a natural setting for studying the one-dimensional Conformal Field Theory (dCFT) living on them. 

A dCFT is characterized by the spectrum of  correlations functions of local and non-local operators. Restricting to line defects, the $n$-point correlation function of a local operator $\cO(s)$ localized on the line  is defined as (an analogous definition holds for the circle)
\begin{equation} \label{eq:corrfunct}
    \llangle \cO(s_n)\cO(s_{n-1})...\cO(s_1) \rrangle = \frac{\langle \text{Tr} W(\infty,s_n) \cO(s_n) W(s_n,s_{n-1})...W(s_2,s_2)\cO(s_1)W(s_1,-\infty) \rangle}{\langle  W(\infty,-\infty) \rangle}\,.
\end{equation}

When we deal with Wilson operators depending on a parameter such as a latitude or a cusp angle, taking $n$ derivatives of the WL with respect to the parameter naturally leads to correlation functions of type \eqref{eq:corrfunct} integrated on the line, where $\cO(s)$ is the operator that appears in the (super)connection $\cL$ multiplied by (a function of) the parameter. 

This fact is at the basis of the proof of the cusp/latitude correspondence \eqref{eq:generalidentity},  giving the remarkable relation between the Bremsstrahlung function as obtained from the cusp anomalous dimension, and the latitude WL. In fact, its proof \cite{Correa:2012at,Correa:2014aga} relies on showing that the double derivative of a latitude WL with respect to the latitude angle and the double derivative of a cusped WL with respect to the cusp angle, which in turn is proportional to the Bremsstrahlung function, give rise to the same two-point function integrated on the circle and on the line, respectively. The key ingredient of the proof is the conformal invariance of the defect, which constraints the form of the two-point functions both on the circle and on the line, and allows to conformally map one into the other.

Already in section \ref{sec:interpolatingBfunction} we gave perturbative evidence that in the case of non-conformal WLs the cusp/latitude correspondence gets spoiled by terms proportional to the conformal anomaly. Here, we provide an alternative interpretation of this result from a defect perspective. We will show that the breaking of the correspondence can be traced back to the appearance of an anomalous dimension for fermionic operators localized on the defect.

To begin with, in section \ref{subsec:thirdwaytoB} we re-compute the Bremsstrahlung function through the evaluation of two-point functions on the defect. In section \ref{sec:corrfunctions} we discuss the general structure of the two-point functions and the emergence of an anomalous dimension for the fermions entering the definition of the defect. Finally, in section \ref{subsec:correspondencerevisited} we show that  this anomalous dimension is indeed responsible for deforming the cusp/latitude correspondence away from the fixed points.

\subsection{A third way to $B$}
\label{subsec:thirdwaytoB}

The starting point for writing the Bremsstrahlung function $B^{\theta}$ in terms of two-point functions is the observation that, regardless of conformality of the defect, our definition \eqref{eq:gammacuspL} for the cusp anomalous dimension together with the RG flow equations lead to the following chain of identities
\begin{equation}
\begin{split}\label{eq:cusp3}
    \Gamma_{\text{cusp}}(\bar\alpha_0,\alpha_0) &\equiv \left( \mu \frac{d}{d\mu}-L \frac{d}{d L} \right) \log Z_{\text{cusp}} = \left( \mu \frac{d}{d\mu}-L \frac{d}{d L} \right) \log \langle W_{\text{cusp}} \rangle_0\\ & =\beta_{g^2}\frac{\partial}{\partial g^2} \log \langle W_{\text{cusp}} \rangle_0\,.
\end{split}
\end{equation}
In the first line we simply replaced $\log Z_{\text{cusp}} = \log\langle W_{\text{cusp}} \rangle_0-\log \langle W_{\text{cusp}} \rangle$ and used the Callan-Symanzik equations for $\langle W_{\text{cusp}} \rangle$. The second line follows from the fact that the explicit dependence of $\langle W_{\text{cusp}} \rangle_0$ on $\mu$ always enters through the combination $\mu L$, whereas an implicit dependence comes through $g^2$.

Using the standard definition $B = \frac{1}{2} \frac{\partial^2}{\partial\theta^2} \Gamma_{\text{cusp}}\big|_{\theta=0}$ together with \eqref{eq:cusp3}, we can write
\begin{equation}\label{eq:bremvia2pt}
    B^{\theta}(\bar\alpha_0,\alpha_0) = \beta_{g^2}\frac{\partial}{\partial g^2} \left( \frac{1}{2}\frac{\partial^2}{\partial \theta^2} \log \langle W_{\text{cusp}} \rangle_0\bigg|_{\theta=0} \right)\,.
\end{equation}
The actual expression for $B^{\theta}$ is eventually obtained by replacing the bare parameters with their renormalized counterparts.

To evaluate explicitly the right hand side of \eqref{eq:bremvia2pt} we start applying $\tfrac12 \frac{\partial^2}{\partial\theta^2}$ to the cusped Wilson line constructed in section \ref{sec:Cusp}.
Setting the geometrical angle $\varphi=0$, we can expand the superconnection \eqref{eq:lineWL} in powers of $\theta$
\begin{equation} \label{eq:operator1}
    \cL = \cL_{\theta=0} + \theta \cL^{(1)}+ \cO(\theta^2)\,,
\qquad \quad 
    \cL^{(1)} = \begin{pmatrix}
        -ig^2\left(O + \bar O  \right) &  \frac{g}{\sqrt{2}} \bar \chi \\ \frac{g}{\sqrt{2}} \chi & -ig^2\left( \hat O +\bar{\hat O}  \right)
    \end{pmatrix}
\end{equation}
where 
\begin{equation}
\label{eq:operators}
\begin{split}
   O = C_3 \bar C^1, \quad  \bar O = \bar{C}^3 C_1, \quad \hat O = \bar C^1 C_3 \,, \quad \bar{\hat{O}} = C_1 \bar C^3 \,,\quad \bar \chi = -i\bar\alpha_0 \sqrt{2}\eta \bar\psi^3 \,,\quad \chi = \alpha_0\sqrt{2} \psi_3 \bar\eta\,.
\end{split}
\end{equation}
It follows that taking the double derivative with respect to $\theta$ gives rise to the $\cL^{(1)}$ two-point function integrated along the line\footnote{We neglect the one-point function $\llangle \cL^{(2)}\rrangle$ since one-point functions of local operators vanish even away from the RG fixed points. Moreover, in order to avoid clattering, from here on we neglect the subscript $0$ in all VEVs, though they are meant to be expressed in terms of bare parameters.}
\begin{equation}\label{eq:cusp2pt}
\begin{split}
    \frac{1}{2}&\frac{\partial^2}{\partial \theta^2}\log\langle W_{\text{cusp}} \rangle \bigg|_{\theta=0} = -\frac{1}{4N} \int_{-\infty}^{\infty} d s_1 \int_{-\infty}^{s_1} ds_2 \llangle \cL^{(1)}(s_1) \cL^{(1)}(s_2) \rrangle = \\ & \quad=  \frac{1}{4N}\int_{-\infty}^{\infty} d s_1 \int_{-\infty}^{s_1} ds_2 \bigg[ 2g^4\left(\llangle O(s_1) \bar O(s_2) \rrangle + \llangle \hat O(s_1) \bar{\hat O}(s_2) \rrangle\right)- g^2\llangle \bar \chi(s_1) \chi(s_2) \rrangle  \bigg]\,.
\end{split}
\end{equation}
 
The evaluation of this expression at order $g^4$ requires the tree-level two-point function for the scalars and the one-loop one for the fermions.  

The lowest order of the $O$, $\hat O$ two-point functions easily evaluates to\footnote{We include a $\mu$ power coming from the overall $g^4$ in \eqref{eq:cusp2pt}.}
\begin{equation}\label{eq:O1result}
    \llangle O(s) \bar O(0) \rrangle^{(0)} =  \llangle \hat O(s) \bar{\hat O}(0) \rrangle^{(0)} = N^3\frac{\Gamma^2(\frac{1}{2}-\epsilon)}{16\pi^{3-2\epsilon}}\frac{\mu^{4\epsilon}}{s^{2-4\epsilon}}\, ,
\end{equation}
and its integrated version then reads
\begin{equation}\label{eq:integratedO}
    \int_{-L}^{L} ds_1 \int_{-L}^{s_1} ds_2 \ \llangle O(s_1) O(s_2) \rrangle^{(0)} = -\frac{N^3}{64\pi^2}\left[ \frac{1}{\epsilon} + 2\log\left( 4\pi e^{\gamma_E+2} \right)+ 2\log\left( 4\mu^2L^2 \right)  \right]\,.
\end{equation}

The tree-level contribution to the fermionic two-point function is
\begin{equation}\label{eq:xxtree}
    \llangle \chi(s)\bar\chi(0) \rrangle^{(0)} = - \bar\alpha_0\alpha_0N^2\frac{\Gamma\left(\frac{3}{2}-\epsilon\right)}{2\pi^{\frac{5}{2}-\epsilon}} \frac{\mu^{2\epsilon}}{s^{2-2\epsilon}}\,,
\end{equation}

At one-loop we have three sources of contributions, one from fermionic arcs 
\begin{equation}
    \llangle \chi(s) \bar \chi(0) \rrangle^{(1)} \Big|_{\text{arcs}} = \bar\alpha_0^2\alpha_0^2 \frac{g^2N^3}{8\pi^{3-2\epsilon}} 
    \frac{\Gamma\left( \tfrac{1}{2}-\epsilon \right)\Gamma\left( \tfrac{3}{2}-\epsilon \right)}{\epsilon}\left( 2^{2\epsilon}L^{2\epsilon} + 2 s^{2\epsilon} \right) \frac{\mu^{4\epsilon}}{s^{2-2\epsilon}}  \,,
\end{equation}
the second one from fermions-gauge vertices
\begin{equation}
    \llangle \chi(s) \bar \chi(0) \rrangle^{(1)} \Big|_{\text{vertices}} = - \bar\alpha_0\alpha_0\frac{g^2N^3}{4^{1-\epsilon}\pi^{3-2\epsilon}}\Gamma(\tfrac{3}{2}-2\epsilon)\Gamma(\tfrac{1}{2}-\epsilon)\Gamma(\epsilon)\cos(\pi\epsilon)\frac{\mu^{4\epsilon}}{s^{2-4\epsilon}}  \,,
\end{equation}
and the third one coming from the expansion of $W$ at the denominator of \eqref{eq:corrfunct}
\begin{equation}\label{eq:W1l}
    \langle W(L,-L) \rangle_0 = 1 - \bar\alpha_0\alpha_0\frac{g^2N}{4\pi}\left[ \frac{1}{\epsilon} + \log\left( 4\mu^2 L^2 \right) + \log\left( 4\pi e^{\gamma_E} \right) \right]\,.
\end{equation}
Summing up these expressions, integrating them together with the tree-level contribution \eqref{eq:xxtree} and expanding in powers of $\epsilon$ we eventually obtain
\begin{equation}\label{eq:integratedX}
\begin{split}
    \int_{-L}^{L} ds_1 \int_{-L}^{s_1} ds_2 \ \llangle &\chi(s_1) \bar\chi(s_2) \rrangle = \bar\alpha_0\alpha_0 \frac{N^2}{8\pi}\left[ \frac{1}{\epsilon} + \log\left( 4\pi e^{\gamma_E} \right) + \log\left(4 \mu^2 L^2\right)\right]-\bar\alpha^2_0\alpha^2_0 \frac{g^2N^3}{16\pi^2\epsilon} \\ &- \bar\alpha_0\alpha_0(\bar\alpha_0\alpha_0-1)\frac{g^2N^3}{16\pi^2\epsilon}\left[ \frac{1}{2\epsilon} + \log\left( 4\pi e^{\gamma_E} \right) + \log(4\mu^2 L^2) \right] + \cO(\epsilon)\,.
\end{split}
\end{equation}

At this point, expressions \eqref{eq:integratedO} and \eqref{eq:integratedX} can be inserted in \eqref{eq:cusp2pt}, leading to $B^{\theta}$, via \eqref{eq:bremvia2pt}
\begin{equation}
\begin{split}
    B^{\theta}(\bar\alpha_0 \alpha_0) =& \bar\alpha_0\alpha_0 \frac{g^2N}{16\pi}\left[ 1 +\epsilon\log\left(4\pi e^{\gamma_E}\right) +\epsilon\log\left(4 \mu^2 L^2\right)\right] - \bar\alpha_0\alpha_0(\bar\alpha_0\alpha_0-1)\frac{g^4N^2}{32\pi^2\epsilon} \\&- (\bar\alpha^2_0\alpha_0^2-1)\frac{g^4N^2}{16\pi^2}  -\bar\alpha_0\alpha_0(\bar\alpha_0\alpha_0-1)\frac{g^4N^2}{16\pi^2}\left[\log\left(4\pi e^{\gamma_E}\right)+\log\left(4\mu^2 L^2\right)\right]\,.
\end{split}
\end{equation}
This is a UV divergent result which, however, is rendered finite upon renormalization of the bare parameters (see \eqref{eqn:renormalizedalphas}). Eventually, replacing $g^2 \to 2\pi/k$, we find
\begin{equation}\label{eq:res2ptB}
    B^{\theta}(\bar\alpha \alpha) =  \frac{1}{8}\frac{N}{k}\bar\alpha\alpha - \frac{1}{4}\frac{N^2}{k^2} (\bar\alpha^2\alpha^2-1)-\frac{1}{16}\frac{N}{k} \beta_{\bar\alpha\alpha}\log\left( \frac{16\pi e^{\gamma_E}}{c_z} \mu^2L^2\right) + \cO \left( \frac{N^3}{k^3} \right) \,.
\end{equation}
This expression coincides with result \eqref{eq:Bfinal} obtained in section \ref{sec:interpolatingBfunction}, if there we choose the particular scheme $c_J = 16\pi e^{\gamma_E}$.


\subsection{Defect correlation functions and anomalous dimensions} \label{sec:corrfunctions}
Going back to the perturbative evaluation of the two-point function for $\chi$ on the line, if we partially expand in $\epsilon$ equations (\ref{eq:xxtree})-(\ref{eq:W1l}), we can cast it in the following form
\begin{equation}\label{eq:chichi}
    \llangle \chi(s)\bar\chi(0)\rrangle =  c_1\, \frac{\mu^{2\epsilon}}{s^{2-2\epsilon}} +   g^2c_2\,\frac{\mu^{4\epsilon}}{s^{2-4\epsilon}}\,,
\end{equation}
where 
\begin{equation}\label{eq:c1c2}
\begin{split}
    c_1 =& -\bar\alpha_0\alpha_0\frac{N^2}{4\pi}\left[ 1 + \epsilon\log\left( 4\pi e^{\gamma_E-2} \right)\right] + \cO(g^4)  \,,\\ c_2 =  & \bar\alpha^2_0\alpha^2_0\frac{N^3}{4\pi^2}  +\bar\alpha_0\alpha_0(\bar\alpha_0\alpha_0-1)\frac{g^2N^3}{4\pi^2}\left[ \frac{1}{2\epsilon} + \log\left( 
    4\pi e^{\gamma_E-2} \right)  \right] + \cO(g^2) \,.
\end{split}
\end{equation}
Since $c_2$ is $\frac{1}{\epsilon}$-divergent, we further expand \eqref{eq:chichi} for $\epsilon \to 0$, keeping at most $\cO(\epsilon)$ terms. The result reads
\begin{equation}
    \llangle \chi(s)\bar\chi(0)\rrangle = \frac{(c_1+g^2 c_2)}{s^2}\left[ 1 + \epsilon \frac{c_1+2g^2c_2}{c_1+g^2c_2}\log\left(\mu^2 s^2\right)  \right]\,.
\end{equation}
The term proportional to $\log s^2$ signals the appearance of an anomalous dimension for $\chi$, which up to $\cO(g^2)$ is given by
\begin{equation}\label{eq:gammaX}
    \gamma_{\chi} = -\epsilon \left( 1 + g^2\frac{c_2}{c_1} \right)= \frac{N}{k}(\bar\alpha\alpha-1)\,.
\end{equation}
 
This result is consistent with the fact that in the $1/2$ BPS case ($\bar\alpha\alpha=1$) $\chi$ is a protected operator, being part of the displacement supermultiplet. For any  other value $\bar\alpha\alpha\ne 1$ it no longer belongs to a protected multiplet and in fact $\gamma_\chi\neq 0$. 

The same investigation can be carried on for the scalar operators $O$ and $\hat O$ on the line. Since they are protected in both the $1/2$ BPS and the $1/6$ BPS bosonic defects, we expect possible anomalous dimension contributions to be proportional to the $\beta$-functions. However, up to one loop it is easy to check that there are no corrections of this type. 

\vskip 10pt

What we have discussed so far holds for the line defect. The question is whether we find a similar pattern on the circle, in particular if we find a similar structure for the two-point functions. In the absence of conformal symmetry we can no longer rely on the line-to-circle mapping to infer the structure of the correlators on the circle from the ones on the line. Thus, in principle, we should re-evaluate correlation functions directly on the circle. 
However, we can by-pass this step by exploiting the results that we have found for the latitude WL, as follows. 

We start from the observation that, taking the derivative of the logarithm of the latitude WL with respect to $\nu$ gives rise to same linear combination of two-point functions as the ones in \eqref{eq:cusp2pt} coming from cusp derivatives. Precisely, 
\begin{equation}\label{eq:latitude2pt}
\begin{split}
    \frac{\partial}{\partial \nu}\log\langle W_{\nu} \rangle\bigg|_{\nu=1} = \frac{1}{N} \int_0^{2\pi}d\tau_1 \int_0^{\tau_1} d\tau_2 \cos\tau_{12}\ \bigg[& 2g^4\left(\llangle O(\tau_1) \bar O(\tau_2) \rrangle + \llangle \hat O(\tau_1) \bar{\hat O}(\tau_2) \rrangle\right)\\ &- g^2\llangle \bar \chi(\tau_1) \chi(\tau_2) \rrangle  \bigg]\,,
\end{split}
\end{equation}
where now the correlation functions are integrated on the maximal circle ($\nu=1$).

We now assume that the two-point functions on the circle are still given by \eqref{eq:chichi}, with the same $c_1, c_2$ coefficients and the obvious replacement $s_{12} \to 2 \sin\frac{\tau_{12}}{2}$.
In principle, this assumption is not supported by conformal invariance and might be spoiled by non-trivial contributions proportional to the $\beta$-functions. However, plugging expressions (\ref{eq:chichi})-(\ref{eq:c1c2}) in \eqref{eq:latitude2pt} and solving the integrals\footnote{The integrals on the circle can be easily performed following \cite{Bianchi:2013rma}.} we find perfect matching with the two-loop expression on the left hand side as read from \eqref{eq:bremsstrahlung_latitude1}.
This is clear evidence that up to $\cO(g^4)$ $c_1$ and $c_2$, and consequently the anomalous dimension in \eqref{eq:gammaX}, are the same on the line and the circle, regardless of conformal invariance.


\subsection{Interpolating cusp/latitude correspondence}
\label{subsec:correspondencerevisited}

Supported by the evidence that \eqref{eq:chichi} holds for both the line and the circle, we now compute explicitly its integrated version on a linear and a circular contour.  For the former, we introduce the usual IR cut-off $L$ and obtain
\begin{equation}
\begin{split}
    \int_{-L}^L ds_1\int_{-L}^{s_1} ds_2 \ \frac{\mu^{2\epsilon}}{s_{12}^{2-2\epsilon}} &= - \frac{1}{2}\left[\frac{1}{\epsilon}+\log(4e^2\mu^2L^2)\right]\,, \\
    \int_{-L}^L ds_1\int_{-L}^{s_1} ds_2 \ \frac{\mu^{4\epsilon}}{s_{12}^{2-4\epsilon}} &= - \frac{1}{2}\left[\frac{1}{2\epsilon}+\log(4e^2\mu^2L^2)\right] \,.
\end{split}
\end{equation}
For the latter, the regulator is given as usual by the radius $R$ of the circle and we find 
\begin{equation}
\begin{split}
    \int_0^{2\pi}d\tau_1 \int_0^{\tau_1}d\tau_2 
    \ \cos\tau_{12}\frac{\mu^{2\epsilon}}{(2\sin\frac{\tau_{12}}{2})^{2-2\epsilon}} &= \pi^2 \left[1+\epsilon\log(\mu^2R^2)\right]\,, \\
    \int_0^{2\pi}d\tau_1\int_0^{\tau_1}d\tau_2  \ \cos\tau_{12}\frac{\mu^{4\epsilon}}{(2\sin\frac{\tau_{12}}{2})^{2-4\epsilon}} &= \pi^2 \left[1+2\epsilon\log(\mu^2R^2)\right] \,.
\end{split}
\end{equation}

Collecting the results on the line, from equation \eqref{eq:cusp2pt} we find
\begin{equation}
\begin{split}
        \beta_{g^2}\frac{\partial}{\partial g^2}\left(\frac12 \frac{\partial^2}{\partial\theta^2}\log \langle W_{\rm cusp}\rangle\Big\vert_{\theta=0}\right)=  \frac{g^2}{4N} \bigg[ c_1 \left(1+\epsilon\log(4e^2\mu^2L^2)\right) +g^2 c_2 \left(1+2\epsilon\log(4e^2\mu^2L^2)\right)\bigg]\,,
\end{split}
\end{equation}
while from equation \eqref{eq:latitude2pt}, on the circle we obtain
\begin{equation}
    \frac{1}{4\pi^2}\frac{\partial}{\partial\nu}\log \langle W_\nu\rangle \bigg|_{\nu=1} = \frac{g^2}{4N} \bigg[ c_1\left(1+\epsilon\log(\mu^2R^2)\right)+ g^2 c_2 \left(1+2\epsilon\log(\mu^2R^2)\right)\bigg]\,.
\end{equation}
The ratio of the two expressions above then reads
\begin{equation}\label{eq:correspondenceBL}
    \frac{\beta_{g^2}\dfrac{\partial}{\partial g^2}\left(\dfrac12 \dfrac{\partial^2}{\partial\theta^2}\log \langle W_{\rm cusp}\rangle\Big|_{\theta=0}\right)}{\dfrac{1}{4\pi^2}\dfrac{\partial}{\partial\nu}\log \langle W_\nu\rangle\Big|_{\nu=1}} = 1 + \epsilon\left( 1 +g^2\frac{c_2}{c_1} \right)\log\left( 4e^2\frac{L^2}{R^2} \right) \,.
\end{equation}
Recalling identity \eqref{eq:bremvia2pt} for the Bremsstrahlung function, the interpolating cusp/latitude correspondence finally reads
\begin{equation}\label{eq:cuspLatitudeNew}
    B^{\theta}(\bar\alpha\alpha) = \left[ 1 -\gamma_\chi \log\left(4e^2\frac{L^2}{R^2}\right) \right]\dfrac{1}{4\pi^2}\dfrac{\partial}{\partial\nu}\log\langle W_\nu\rangle\bigg|_{\nu=1}\,,
\end{equation}
where in \eqref{eq:cuspLatitudeNew} we have recognized the $\chi$ anomalous dimension as given in \eqref{eq:gammaX}.

In conclusion, away from the fixed points the terms spoiling the usual cusp/latitude correspondence can be traced back to the anomalous dimension acquired by fermions localized on the defect. 

We stress that this result is valid perturbatively, up to two loops. At this order the deforming $\gamma_\chi$ term enters multiplied by the one-loop contribution to the $\nu$-derivative of the latitude. Since this is proportional to $\bar\alpha \alpha$, we easily reconstruct the one-loop $\beta$-function $\beta_{\bar\alpha \alpha}$. Therefore, identity \eqref{eq:cuspLatitudeNew} is in perfect agreement with \eqref{eq:Blatitude} in the $c_J = 16 \pi e^{\gamma_E}$ scheme, consistently with what we have found in \eqref{eq:res2ptB}. 

It is interesting to note that factor appearing in front of the $\nu$-derivative in \eqref{eq:cuspLatitudeNew} is scheme independent. It depends only 
 on the scales of the linear and circular contours. Of course, scheme dependence in \eqref{eq:cuspLatitudeNew} is still present, being encoded in the explicit expressions of $ B^{\theta}$ and $\partial_\nu \log\langle W_\nu\rangle$.

\vskip 40pt 

\section*{Acknowledgements}

LC, SP and MT were partially supported by the INFN grant {\it Gauge Theories, Strings and Supergravity (GSS)}.
DT is supported in part by the INFN grant {\it Gauge and String Theory (GAST)}. DT would like to thank FAPESP’s partial support through the grants 2016/01343-7 and 2019/21281-4.

\appendix

\section{Conventions and Feynman rules}
\label{app:abjm}

We follow the conventions in \cite{Bianchi:2014laa}. We work in three-dimensional Euclidean space with coordinates $x^{\mu}=(x^0,x^1,x^2)$. The three-dimensional gamma matrices are defined as
\begin{equation}\label{eq:gamma}
    (\gamma^{\mu})^{ \ \beta}_{ \alpha}=(-\sigma^3,\sigma^1,\sigma^2)_\alpha^{\ \beta}\,,
\end{equation}
with $(\sigma^{i} )^{ \ \beta}_{  \alpha}$ ($\alpha,\beta=1,2$) being the Pauli matrices, such that $\gamma^{\mu}\gamma^{\nu}=\delta^{\mu\nu}+i\epsilon^{\mu\nu\rho}\gamma_{\rho}$, where $\epsilon^{123}=\epsilon_{123}=1$ is totally antisymmetric. Spinorial indices are lowered and raised as $(\gamma^{\mu})^{\alpha}_{\ \beta}=\epsilon^{\alpha\gamma}(\gamma^{\mu})^{\ \delta}_{\gamma} \epsilon_{\beta\delta}$, with $\epsilon_{12}=-\epsilon^{12}=1$. The Euclidean action of $U(N)_k\times U(N)_{-k}$ ABJM theory is
\begin{equation}\label{eq:ABJMaction}
\begin{split}
    S_{\textrm{ABJM}}=&\frac{k}{4\pi} \int d^3 x\,  \epsilon^{\mu\nu\rho}\Big\{ -i\text{Tr}\left( A_{\mu}\partial_{\nu}A_{\rho} +\frac{2i}{3}A_{\mu}A_{\nu}A_{\rho} \right)+i\text{Tr}\left( \hat A_{\mu}\partial_{\nu}\hat A_{\rho} +\frac{2i}{3}\hat A_{\mu}\hat A_{\nu}\hat A_{\rho} \right) \\ & + \text{Tr}\left[ \frac{1}{\xi}(\partial_{\mu}A^{\mu})^2 - \frac{1}{\xi}(\partial_{\mu}\hat A^{\mu})^2 +\partial_{\mu} \bar c D^{\mu}c-\partial_{\mu} \bar{\hat c}D^{\mu}\hat c\right] \Big\} \\ & + \int d^3x \text{Tr}\left[ D_{\mu} C_I D^{\mu} \bar C^I +i\bar\psi^I \gamma^{\mu}D_{\mu} \psi_I \right]\\ & \begin{split}- \frac{2\pi i}{k}\int d^3 x \text{Tr}\Big[& \bar C^I C_I \psi_J \bar\psi^J - C_I \bar C^I \bar\psi^J \psi_J + 2C_I\bar C^J \bar\psi^I\psi_J \\ &-2\bar C^I C_J \psi_I \bar\psi^J - \epsilon_{IJKL}\bar C^I \bar \psi^J \bar C^K \bar \psi^L +\epsilon^{IJKL} C_I \psi_J C_K \psi_L \Big]+ S^{\text{bos}}_{\text{int}}\,,\end{split}
\end{split}
\end{equation}
where $A_\mu$ and $\hat{A}_\mu$ are the connections of the two gauge groups, whereas $C_I$ and $\psi_I$ describe scalar and fermion matter, respectively. The covariant derivatives are defined as
\begin{equation}
\begin{split}
    &D_{\mu} C_I = \partial_{\mu} C_I +i A_{\mu} C_I -i C_I \hat A_{\mu}\,, \qquad D_{\mu} \bar C^I=\partial_{\mu} \bar C^I -i \bar C^I A_{\mu} + i\hat A_{\mu} \bar C^I\,, \\ & D_{\mu}\bar\psi^I=\partial_{\mu} \bar\psi^I + iA_{\mu} \bar\psi^I - i\bar\psi^I \hat A_{\mu}\,, \qquad D_{\mu} \psi_I =\partial_{\mu}\psi_I -i\psi_I A_{\mu} +i\hat A_{\mu} \psi_I\,.
\end{split}
\end{equation}

We work in Landau gauge for vector fields and in dimensional regularization with $d=3-2\epsilon$. The tree-level propagators are (with $g^2=2\pi/k$)
\begin{equation}
\label{eqn:propagator}
    \begin{split}
        \langle (A_{\mu})_p^{\ q}(x)(A_{\nu})_r^{\ s}(y)\rangle^{(0)} &=\delta_p^s\delta_r^q\, i g^2 \, \frac{\Gamma(\frac{3}{2}-\epsilon)}{2\pi^{\frac{3}{2}-\epsilon}}\frac{\epsilon_{\mu\nu\rho}(x-y)^{\rho}}{|x-y|^{3-2\epsilon}},\\
        \langle (\hat A_{\mu})_{\hat p}^{\ \hat q}(x)(\hat A_{\nu})_{\hat r}^{\ \hat s}(y)\rangle^{(0)} &=-\delta_{\hat p}^{\hat s}\delta_{\hat r}^{\hat q} \, i g^2 \, \frac{\Gamma(\frac{3}{2}-\epsilon)}{2\pi^{\frac{3}{2}-\epsilon}}\frac{\epsilon_{\mu\nu\rho}(x-y)^{\rho}}{|x-y|^{3-2\epsilon}},\\
        \langle  (\psi_I^{\alpha})_{\hat i}^j(x) (\bar\psi_{\beta}^J)_k^{\hat l} (y) \rangle^{(0)} & = -i\delta_I^J\delta_i^{\hat l}\delta_k^{j } \frac{\Gamma(\frac{3}{2}-\epsilon)}{2\pi^{\frac{3}{2}-\epsilon}}\frac{(\gamma_{\mu})^{\alpha}_{\ \beta}(x-y)^{\mu}}{|x-y|^{3-2\epsilon}}\\ & =i\delta_I^J\delta_i^{\hat l}\delta_k^{j } (\gamma_{\mu})^{\alpha}_{\ \beta}\partial_{\mu}\left( \frac{\Gamma(\frac{1}{2}-\epsilon)}{4\pi^{\frac{3}{2}-\epsilon}} \frac{1}{|x-y|^{1-2\epsilon}}\right), \\
        \langle (C_I)_i^{\hat j}(x)(\bar C^J)_{\hat k}^l(y)\rangle^{(0)} &= \delta_I^J \delta_i^l \delta_{\hat k}^{\hat j}  \frac{\Gamma(\frac{1}{2}-\epsilon)}{4\pi^{\frac{3}{2}-\epsilon}}\frac{1}{|x-y|^{1-2\epsilon}} ,
    \end{split}
\end{equation}
while the one-loop propagators read 
\begin{equation}
\label{eqn:onelooppropagator}
\begin{split}
    \langle (A_{\mu})_p^{\ q}(x) (A_{\nu})_r^{s}(y) \rangle^{(1)} &= \delta_p^s\delta_r^q \, g^4 N \, \frac{\Gamma^2(\frac{1}{2}-\epsilon)}{4\pi^{3-2\epsilon}}\left[ \frac{\delta_{\mu\nu}}{|x-y|^{2-4\epsilon}}-\partial_{\mu}\partial_{\nu}\frac{|x-y|^{2\epsilon}}{4\epsilon(1+2\epsilon)} \right] ,\\
    \langle (\hat A_{\mu})_{\hat p}^{\ \hat q}(x) (\hat A_{\nu})_{\hat r}^{ \hat s}(y) \rangle^{(1)} &= \delta_{\hat p}^{\hat s}\delta_{\hat r}^{\hat q} \, g^4 N \,\frac{\Gamma^2(\frac{1}{2}-\epsilon)}{4\pi^{3-2\epsilon}}\left[ \frac{\delta_{\mu\nu}}{|x-y|^{2-4\epsilon}}-\partial_{\mu}\partial_{\nu}\frac{|x-y|^{2\epsilon}}{4\epsilon(1+2\epsilon)} \right] .    
\end{split}
\end{equation}
The latin indices are color indices. For instance, $(A_{\mu})_p^{\ q} \equiv A_\mu^a (T^a)_p^{\ q}$ where $T^a$ are $U(N)$ generators in fundamental representation.

We work in the large $N, k$ limit with $g^2 N \ll 1$. 


\bibliographystyle{utphys2}
\bibliography{refs}

\end{document}